\documentclass[11pt]{article}
\usepackage[letterpaper, left=1.25in, right=1.25in, top=1.25in, bottom=1.25in]{geometry}
\usepackage[rm,bf]{titlesec}
\usepackage{enumitem}
	
\usepackage{amsmath}
\usepackage{amsthm}
\usepackage{multirow}
\usepackage{booktabs}
\usepackage{graphicx}
\usepackage{amssymb}
\usepackage{amsfonts}
\usepackage{amsmath,amsthm}
\usepackage[footnotesize,bf]{caption}
\usepackage{natbib}
\usepackage{setspace}
\usepackage[dvipsnames,usenames]{color}
\usepackage{pstricks,pstricks-add}
\usepackage{url}
\usepackage[usenames]{color}
\usepackage[breaklinks, colorlinks, citecolor=blue, linkcolor=blue, urlcolor=blue]{hyperref}
\usepackage{lscape}
\usepackage{rotating}

\newtheorem{theorem}{Theorem}

\newtheorem{theorem*}[theorem]{Theorem*}

\newtheorem{definition}{Definition}
\newtheorem{definition*}[definition]{Definition*}

\newtheorem{example*}[example]{Example*}

\newtheorem{lemma*}[lemma]{Lemma*}

\newtheorem{proposition}{Proposition}
\newtheorem{proposition*}[proposition]{Proposition*}
\newtheorem{remark}{Remark}

\newcommand{\Bcal}{\mathcal{B}}

\newcommand{\N}{\mathbb{N}}

\newpsobject{grilla}{psgrid}{subgriddiv=1,griddots=10,gridlabels=6pt}

\newcommand{\vfivefive}{\rule{0pt}{5.25ex}}

\begin{document}
\bibliographystyle{elsart-harv}
\title{Empirical bias and efficiency of alpha-auctions: experimental evidence\footnote{The experiments in this paper were funded with the financial support of the Texas A\&M Humanities and Social Science Enhancement of Research Capacity Program and conducted by Daniel Stephenson, Raisa Sara, and Manuel Hoffman. This paper benefited from the comments of audiences at U. of Bristol, Caltech, Chapman U., UC Santa Barbara, U. Maryland, U. of Southampton, U. of Virginia, the 2016 World and North American Economic Science Association meetings, and TETC17.  All errors are our own.}}
\date{\today}

\author{Alexander L. Brown\thanks{
 \href{mailto:alexbrown@tamu.edu}{alexbrown@tamu.edu}; \href{http://people.tamu.edu/\%7Ealexbrown}{http://people.tamu.edu/$\sim$alexbrown}} \ \ and Rodrigo A. Velez\thanks{
\href{mailto:rvelezca@tamu.edu}{rvelezca@tamu.edu}; \href{https://sites.google.com/site/rodrigoavelezswebpage/home}{https://sites.google.com/site/rodrigoavelezswebpage/home}} \\\small{\textit{Department of
Economics, Texas A\&M University, College Station, TX 77843}} }
\maketitle

\begin{abstract}
We experimentally evaluate the comparative performance of the winner-bid, average-bid, and loser-bid auctions for the dissolution of a partnership. The analysis of these auctions based on the empirical equilibrium refinement of \citet{Velez-Brown-2019-SP} reveals that as long as behavior satisfies weak payoff monotonicity, winner-bid and loser-bid auctions necessarily exhibit a form of bias when empirical distributions of play approximate best responses \citep{Velez-Brown-2019-EBA}. We find support for both weak payoff monotonicity and the form of bias predicted by the theory for these two auctions. Consistently with the theory, the average-bid auction does not exhibit this form of bias. It has lower efficiency that the winner-bid auction, however.
\medskip

\begin{singlespace}
\medskip

\end{singlespace}
\end{abstract}

\newpage

\section{Introduction}\label{Sec:intro}

The average-bid auction (AB) is an intuitive mechanism for the dissolution of a partnership, i.e., the allocation of an indivisible good among two agents when monetary compensation is possible. Fair Outcomes Inc., a web service designed by some of the leaders in fair allocation theory uses this protocol for their  ``fair buy-sell'' service, which they describe as follows:

\begin{quotation}\textit{
``The system provides each party with an opportunity to enter data into the system under an escrow arrangement, including confidential data specifying a monetary value for the property at which that party would be willing either to sell its share to the other side or buy out the other side's share (similar to the proposal made by the initiating party under a traditional ``buy-sell'' arrangement). The system compares the values entered by the parties and announces a sale to the party that specified the higher value. However, the sale price is set at the midpoint between the two values.''} \url{https://www.fairoutcomes.com/fb.html}\linebreak(retrieved on May 3, 2019).
\end{quotation}

AB makes part of a family of mechanisms that can be described as follows. The arbitrator asks agents to bid for the good. Then assigns the good to a higher bidder. This agent pays to the other agent a convex combination of both bids with fixed weights multiplying the winner and loser bids. AB determines the price with equal weights for both bids. Alternative intuitive and practical price determination rules are to set the price as the winner bid (WB auction), or the loser bid (LB auction).

Under independent symmetric private values, all mechanisms in this family have efficient Bayesian Nash equilibria  \citep{Cramton-et-al-1987-Eca}. When information is complete they obtain in Nash equilibria only efficient allocations that would result from a market with equal incomes, i.e., they induce intuitively equitable outcomes \citep{Velez-Brown-2019-EBA}. Interestingly, with complete information WB and LB essentially have the same Nash equilibrium outcomes \citep{Velez-Brown-2019-EBA}.

The purpose of this paper is to experimentally evaluate the comparative performance of LB, AB, and WB in a complete information environment. This is a relevant benchmark for an arbitrator who oversees a divorce settlement or the dissolution of a long standing partnership. Our main achievement is to document differences in agents' behavior in these mechanisms that are not predicted by Nash equilibrium, but are in line with the theoretical results produced by the recently introduced empirical equilibrium analysis of \citet{Velez-Brown-2019-SP} for these auctions. Our work constitutes the first experimental study that tests the foundations of this theory and the comparative statics it predicts. Indeed, we organize our presentation around these two objectives.

Empirical equilibrium is defined by means of the following thought experiment. Consider a researcher who samples behavior in normal-form games and constructs a theory that explains this behavior. The researcher determines the plausibility of Nash equilibria based on the empirical content of the theory by requiring that Nash equilibria be in its closure. That is, if a Nash equilibrium cannot be approximated to an arbitrary degree by the empirical content of the researcher's theory, it is identified as implausible or unlikely to be observed. Empirical equilibrium is the refinement so defined based on the non-parametric theory that each agent chooses actions with higher probability only when they are better for her given what the other agents are doing. More precisely, an empirical equilibrium is a Nash equilibrium that is the limit of a sequence of behavior satisfying \emph{weak payoff monotonicity}, i.e., between two alternative actions for an agent, say $a$ and $b$, if the agent plays $a$ with higher frequency than $b$, it is because given what the other agents are doing, $a$ has higher expected utility than $b$. Remarkably, this property is a common factor in some of the most popular models that have been proposed to account for observed behavior in experiments.\footnote{These models include the exchangeable randomly perturbed payoff models \citep{Harsanyi-1973-IJGT,VanDamme-1991-Springer}, the control cost model \citep{VanDamme-1991-Springer},  the structural QRE model \citep{mckelvey:95geb},  the regular QRE models \citep{Mackelvey-Palfrey-1996-JER,Goeree-Holt-Palfrey-2005-EE}. The common characteristic of these models that make them suitable the purpose of equilibrium refinement is that their parametric forms are independent of the game in which they are applied, and have been successful in replicating comparative statics in a diversity of games \citep{Goeree-Holt-Palfrey-2005-EE}.}

One can give  empirical equilibrium a static or dynamic interpretation. First, its definition simply articulates the logical implication of the hypothesis that the researcher's theory is well specified, for this hypothesis is refuted by the observation of a Nash equilibrium that is not an empirical equilibrium. Alternatively, suppose that the researcher hypothesizes that behavior will converge to a Nash equilibrium through an unmodeled evolutionary process that produces a path of behavior that is consistent with her theory. Then, the researcher can also conclude that the only Nash equilibria that will be approximated by behavior are empirical equilibria.

The reality is that Nash equilibria are usually not observed in laboratory experiments. Thus, empirical equilibrium per se is not the most appropriate theoretical benchmark to analyze data from experiments. However, this refinement, because of its definition based on proximity of empirically plausible behavior, produces two sets of predictions that are plausible to be verified in experiments. First, if a Nash equilibrium is not empirical, it is unlikely to be observed, even approximately. Second, if weak payoff monotonicity is plausible and agents frequencies of play evolve towards mutual best responses, behavior will eventually move towards an empirical equilibrium. For WB and LB this analysis produces sharp comparative statics.

We assume that agents' have quasi-linear preferences with values $v_l<v_h$, i.e., there is a higher valuation (HV) and a lower valuation (LV) agent. Under this assumption, the winning bids in the Nash equilibria (including equilibria in mixed strategies) in both WB and LB belong to the set $\{v_l/2,....,v_h/2\}$ \citep{Velez-Brown-2019-EBA}. All Nash equilibria of WB and LB are essentially efficient.\footnote{When the arbitrator breaks ties uniformly at random, there may be some Nash equilibria of WB and LB in which the LV agent wins the good with positive probability. This probability goes to zero when the number of bids in the Nash range is increases. With the tie breaker in our experiments all Nash equilibria of WB and LB are efficient.} Thus, between the Nash equilibria associated with two bids in the Nash range, the LV agent prefers the right one (being paid more) and the HV agent prefers the left one (to pay less).

Empirical equilibrium reveals sharp differences between WB, AB, and LB \citep{Velez-Brown-2019-EBA}. The empirical equilibria of WB have winning bids only on the left half of the Nash range. The empirical equilibria of LB have winning bids only on the right half of the Nash range. The exact bids that can be sustained in each case depend on the tie breaker that is used in the auction and on the number of bids that are available outside the Nash range. In our experiments we implement a deterministic tie breaker that favors the HV agent. When there are enough bids outside the Nash range, the empirical equilibria of WB are sustained by winning bids only on the left third of the Nash range, and the empirical equilibria of LB are sustained by winning bids only on the right third of the Nash range (Theorems~\ref{Thm:characterization-ee-WB} and~\ref{Thm:characterization-ee-LB}).\footnote{\citet{Velez-Brown-2019-EBA} provide a complete characterization of the empirical equilibria of WB and LB for the symmetric tie breaker. The result is qualitatively the same as with the asymmetric tie breaker that we use in our experiments (see Sec.~\ref{Sec:EEtheory}).} Thus, empirical equilibrium analysis predicts that if empirical distributions of play in WB and LB are weakly payoff monotone, they will exhibit opposed forms of bias when these distributions approach mutual best responses. WB favors the HV agent and LB favors the LV agent. Moreover, this analysis produces comparative statics for bid distributions both between LV and HV agents in the same auction, and for all types between two different auctions. For AB, the whole range of assignments in the Nash range are outcomes of empirical equilibria. Thus, this theory does not predict a necessary bias of this mechanism.

The comparative statics predicted for payoffs and average bids by empirical equilibrium analysis for WB and LB are supported by data (Sec.~\ref{Sec:mechanism-bias}-\ref{Sec:bids}). WB is biased in favor of the HV agents and LB is biased towards LV agents. AB is not necessarily biased in favor of any of these populations. Even though effects like rounding induce violations of weak payoff monotonicity in our data, there is support for a positive association of bids with their expected utility (Sec.~\ref{Sec:payoff-monotonicity}). We can conclude that the bias of WB and LB is likely to be a structural invariant of these auctions. Thus, AB is indeed on a sweet spot that may balance better the incentives of these auctions with respect to the equity that is obtained. The ranking of auctions in terms of efficiency favors WB over the other two formats, and if anything favors AB over LB (Sec.~\ref{Sec:efficiency}). Thus, we document a tradeoff between efficiency and equity of these mechanisms.

There is little experimental evidence on the comparative performance of WB, AB, and LB. \citet{Kittsteiner-Ockenfels-Trhal-2012} and \citet{Brown-Velez-2016-GEB} evaluate WB under incomplete and complete information structures, respectively, and compare it with an alternative sequential mechanism in which one agent proposes a price and then the other decides either to buy or sell at that price. \citet{Brown-Velez-2016-GEB} documented agents' propensity to underbid compared to their Nash equilibrium predictions in WB, which is replicated in our experiments. This study also observed that this pattern of behavior is replicated by the QRE models of \citet{mckelvey:95geb}. Following this lead \citet{Velez-Brown-2019-EBA} defined empirical equilibrium and applied this refinement for the study of strategy-proof mechanisms, and \citet{Velez-Brown-2019-EBA} characterized the whole set of empirical equilibria of WB and LB. Our paper complements these two studies by providing experimental evidence supporting the foundations of this theory and the comparative statics it produces.

As part of the empirical equilibrium agenda, our paper has similarities in spirit with the literature that analyses mechanisms for populations of (as if) boundedly rational agents \citep[c.f.,][]{Anderson-et-al-1998-JPE,ANDERSON-et-al-2001-GEB,Eliaz-2002-RStud,
de-Clippel-2014-AER,MASUDA-et-al-2014-GEB,de-Clippel-et-al-2017-levelk,Kneeland-2017-levelk} and implementation in equilibria resulting from convergence processes \citep{Cabrales2012,TUMENNASAN-2013-GEB}. The closest among these papers is \citet{MASUDA-et-al-2014-GEB}, which shows how a certain mechanism for the provision of public goods performs well both in experiments and in theory among populations of boundedly rational agents who follow certain parametric deviations from utility maximization.

The remainder of the paper is organized as follows. Sec.~\ref{Sec:model} introduces our partnership dissolution model and develops our theoretical benchmark in detail. Sec.~\ref{Sec:Exp-design} presents our experimental design. Sec.~\ref{Sec:Exp-results} presents our experimental results. Sec.~\ref{Sec:Discussion} concludes and discusses the alternative predictions of undominated equilibria and other refinements that discard all weakly dominated behavior.


\section{Model and theoretical benchmark}\label{Sec:model}

\subsection{Partnership dissolution}

There are two agents $N\equiv\{l,h\}$ who symmetrically own an indivisible good and need to allocate it to one of them. Monetary compensation is possible and no money is burned in the allocation process.  We refer to agents as $i$ and $-i$ whenever it is convenient. Agent~$i$'s utility from  receiving the object and paying $p_i$ to the other agent is $v_i-p_i$; the agent's utility from receiving transfer $p_i$ and no object is $p_i$. We assume agents are expected utility maximizers. For analytical convenience we will assume valuations are even positive numbers and that there is a maximum valuation $\overline{v}$. We also assume $v_l<v_h$ and denote the profile of valuations by $v\equiv(v_l,v_h)$. The set of possible allocations is that in which an agent receives the object and transfers an amount $p\in\Bcal\equiv\{0,1,...,\overline{p}\}$ with $\overline{p}\geq \overline{v}/2$, to the other agent.

We consider the following family of arbitration protocols.

\begin{definition}\rm The \textit{$\alpha$-auction (with tie-breaker $\gamma$)}  is the mechanism in which each agent selects a bid in  $\Bcal$. An agent with the highest bid receives the object. In case of a tie, the HV agent receives the object with probability $\gamma$. The agent who receives the object pays $\alpha$(winner bid)+$(1-\alpha)$(loser bid) to the other agent. The \textit{winner-bid auction} (WB) corresponds to $\alpha=1$; the \textit{average-bid auction} (AB) corresponds to $\alpha=1/2$; the \textit{loser-bid auction} (LB) corresponds to $\alpha=0$. We refer to WB and LB as the \textit{extreme-price} auctions and to all other as \textit{interior-price} auctions.
\end{definition}

To simplify the interface in experiments, we implemented in the laboratory the tie-breaker $\gamma=1$. Our theoretical benchmark for both the symmetric and asymmetric tie-breakers produces the same comparative statics (see Sec.~\ref{Sec:EEtheory} for details). In the remainder of the paper whenever we refer to an $\alpha$-auction the default tie-breaker is $\gamma=1$.

\subsection{Theoretical benchmark}\label{Sec:EEtheory}

We assume that agents know each other's valuations for the object.  With this informational structure each $\alpha$-auction induces a complete information game. For WB, AB, and LB we denote these games by  $WB(v)$, $AB(v)$, and $LB(v)$, respectively. A (mixed) strategy of agent $i$, which we denote by $\sigma_i\in\Delta(\Bcal)$, is a probability distribution on the set of possible bids. Agent~$i$'s expected bid given strategy $\sigma_i$ is $E_{\sigma_i}(b)$. A profile of mixed strategies $\sigma\equiv(\sigma_l,\sigma_h)$ is a \textit{Nash equilibrium of the $\alpha$-auction with valuations $v$} if each agent's strategy places positive probability only on bids that maximize the agent's expected utility given the strategy of the other agent. Agent $i$'s payoff when distributions of play are $\sigma$ is $\pi_i(\sigma)$, where, for convenience, we drop the dependence of this amount on $v$.

To characterize the Nash equilibria of the $\alpha$-auctions, it is useful to define agent~$i$'s net valuation as the amount $c_i\equiv v_i/2$. Note that the agent is indifferent between receiving the object and transferring $c_i$ to the other agent, and receiving no object and being transferred $c_i$. This means that the agent would be indifferent between being the winner or the loser in an $\alpha$-auction when the transfer from the winner to the loser is $c_i$. We  follow \citet{Tadenuma-and-Thomson-1995-TD} and refer to the amount $ES(v)\equiv c_h-c_l$ as the \textit{equity surplus for $v$}.

\begin{proposition}\label{Prop:charc-equili-auctions}\rm \citep{Velez-Brown-2019-EBA}

\begin{enumerate}\item Let $\sigma$ be a Nash equilibrium of an extreme-price auction with valuations $v$. Then, there is $p\in\{c_l,...,c_h\}$ in the support of both $\sigma_l$ and $\sigma_h$ such that the support of $\sigma_{l}$ belongs to $\{0,...,p\}$ and the support of $\sigma_{h}$ belongs to $\{p,...,\overline{p}\}$.

\item The set of Nash equilibrium payoffs of each extreme-price auction for valuations $v$ is $\{(\pi_l,\pi_h):\pi_l=c_l+t,\pi_h=c_h+(ES(v)-t),t=0,...,ES(v)\}$.
\end{enumerate}
\end{proposition}

Nash equilibria of the extreme-price auctions have a simple structure. In each equilibrium the HV agent receives the object and transfers, with certainty, a given amount in $\{c_l,...,c_h\}$. That is, for each Nash equilibrium there is a unique payoff determinant bid  in $\{c_l,...,c_h\}$; and for each bid in this set there is a Nash equilibrium with this payoff determinant bid. Because of this one-to-one relation between equilibrium payoffs and payoff-determinant bids, we refer to $\{c_l,...,c_h\}$ as the Nash range. Note that both extreme-price auctions are equivalent in terms of the Nash equilibrium payoffs they generate.

Observe also that extreme-price auctions produce in Nash equilibria only allocations that maximize aggregate utility (are \textit{efficient}), which span the whole spectrum of the so called \textit{envy-free} set, i.e., the allocations at which no agent prefers the allotment of the other \citep{Foley-1967-YEE}. This set also corresponds to the allocations that can be sustained by prices as competitive equilibrium outcomes in a market with equal incomes, an intuitively equitable institution \cite[see][for an extended discussion]{Brown-Velez-2016-GEB}. At each such allocation agent $l$ receives a transfer greater than or equal to $c_l$ and agent $h$ receives the object and transfers no more than $c_h$. Thus, each of these allocations determines a division of $c_h-c_l$, the equity surplus, between the agents: If transfer is $p$, agent $l$ captures $p-c_l$ and agent $h$ captures $c_h-p$. Thus, the agents have opposite preferences on the Nash range. Between two equilibria, agent $l$ prefers the right one and agent $h$ the left one.

When $\alpha$ is interior, one can easily see that equal bids in the Nash range constitute a \textit{strict Nash equilibrium} of the $\alpha$-auction, i.e., a profile of strategies in which each agent is playing her unique best response to the strategy of the other agent. The next proposition, whose straightforward proof we omit, follows.

\begin{proposition}\label{Prop:strict-equil}\rm For each $v$ the set of strict Nash equilibrium payoffs of each interior-price auction contains those of the extreme price auctions.
\end{proposition}

\citet{Velez-Brown-2018-EE} propose to refine the set of Nash equilibria by means of approachability by behavior satisfying the following regularity property for which there is empirical support in diverse strategic situations \citep[see][]{Goeree-Holt-Palfrey-2016-Book}. We specialize the definition for the particular games we study.

\begin{definition}[\citealp{Velez-Brown-2018-EE}]\rm Let $\sigma\equiv (\sigma_l,\sigma_h)$ be a profile of mixed strategies in an $\alpha$-auction with valuations $v$. Then, $\sigma$ is \textit{weakly payoff-monotone} for $v$ if for each $i\in\{l,h\}$ and each pair of bids $\{b,d\}\subseteq\Bcal$, $\sigma_i(b)>\sigma_i(d)$ implies that the expected payoff of bid $b$ for agent~$i$ in the $\alpha$-auction with valuations $v$ given $\sigma$ is greater than the corresponding expected payoff of bid $d$.
\end{definition}

Weak payoff monotonicity simply requires that differences in behavior reveal differences in expected payoffs. This property is satisfied by the most popular models for the analysis of experimental data, e.g., the control cost models of \cite{VanDamme-1991-Springer}, the monotone structural QRE models of \citet{mckelvey:95geb}, the regular QRE models of \citet{Goeree-Holt-Palfrey-2005-EE}.

\citet{Velez-Brown-2018-EE} define an empirical equilibrium of a game to be one of its Nash equilibria that can be approximated by weakly payoff monotone behavior. Under the hypothesis that behavior is weakly payoff monotone and approximates a Nash equilibrium,  this behavior must approach an empirical equilibrium.

In our partnership dissolution environment, it is technically feasible to characterize the set of empirical equilibria of EPAs \citep{Velez-Brown-2019-EBA}. Essentially, the empirical equilibria of these auctions separates. When there are enough bids outside the Nash range, the empirical equilibria of  WB (with symmetric tie-breaker) are sustained by winning bids on the left fifth of the Nash range. Symmetrically, the empirical equilibria of LB are sustained by winning bids on the right fifth of the Nash range. \citet{Velez-Brown-2019-EBA}'s analysis can be reproduced for the asymmetric tie-breaker that we used in our experiments. The result is preserved qualitatively.  When there are enough bids outside the Nash range, the empirical equilibria of WB (with asymmetric tie-breaker) are sustained by winning bids on the left third of the Nash range. Symmetrically, the empirical equilibria of LB are sustained by winning bids on the right third of the Nash range.

Even though agents behavior in simultaneous games may approximate Nash equilibria, it is usually noisy. Thus, one can expect that the hypothesis that behavior conforms to an empirical equilibrium will be easily rejected from data. Because of this, for the purpose of generating meaningful comparative statics to test experimentally, one should look into the comparative statics that convergence to an empirical equilibrium produce. That is, one should look into the characteristics of behavior that must be observed when this behavior reaches a ``reasonable'' proximity to the empirical equilibria of the game.

\begin{theorem}\label{Thm:characterization-ee-WB}\rm Let $\sigma$ be a Nash equilibrium of $WB(v)$ and $\{\sigma^\lambda\}_{\lambda\in\N}$ a sequence of weakly-monotone distributions for $WB(v)$ such that as $\lambda\rightarrow\infty$, $\sigma^\lambda\rightarrow\sigma$. Let $t(v)\equiv\max\{2c_h/3-c_l,ES(v)/3\}+1$.  Then, there exists $\Lambda\in\N$ such that for each $\lambda\geq \Lambda$,
      \begin{enumerate}
        \item If $v_l\geq v_h/3$, $E_{\sigma_l^\lambda}(b)< c_l+1$;
        \item $c_l-1<E_{\sigma_h^\lambda}(b)<c_l+t(v)$;
        \item $\pi_l(\sigma^\lambda)<c_l+t(v)$ and $\pi_h(\sigma^\lambda)>c_h+(ES(v)-t(v))$;
        \item If $\pi_l(\sigma)>c_l$, $E_{\sigma_l^\lambda}(b)<E_{\sigma_h^\lambda}(b)$.
      \end{enumerate}
\end{theorem}

\begin{theorem}\label{Thm:characterization-ee-LB}\rm Let $\sigma$ be a Nash equilibrium of $LB(v)$ and $\{\sigma^\lambda\}_{\lambda\in\N}$ a sequence of weakly-monotone distributions for $LB(v)$ such that as $\lambda\rightarrow\infty$, $\sigma^\lambda\rightarrow\sigma$. Let $t(v)\equiv\max\{2(\overline{p}-c_l)/3-c_h,ES(v)/3\}+1$. Then, there exists $\Lambda\in\N$ such that for each $\lambda\geq \Lambda$,
      \begin{enumerate}
        \item If $c_h\leq \overline{p}-(\overline{p}-c_l)/3$, $E_{\sigma_h^\lambda}(b)\geq c_h-1$;
        \item $c_h-t(v)<E_{\sigma_l^\lambda}(b)<c_h+1$;
        \item $\pi_h(\sigma^\lambda)<c_h+t(v)$ and $\pi_l(\sigma^\lambda)>c_l+(ES(v)-t(v))$.
        \item If $\pi_h(\sigma)<c_h$, $E_{\sigma_h^\lambda}(b)>E_{\sigma_l^\lambda}(b)$.
      \end{enumerate}
\end{theorem}

Since the proofs of Theorems~\ref{Thm:characterization-ee-WB} and~\ref{Thm:characterization-ee-LB} follow from modifications of the main arguments in \citet{Velez-Brown-2019-EBA}, we present them in an Online Appendix.

\begin{figure}[h]
\centering
\begin{pspicture}(-1,-1.7)(11,2)
\psline{<->}(-.5,0)(10.5,0)
\psdots[dotsize=5pt,dotstyle=|](0,0)(10,0)
\rput[c](0,-.28){$\mbox{\footnotesize$0$}$}
\rput[c](10,-.28){$\mbox{\footnotesize$\overline{p}$}$}
\psdots[dotsize=4pt,dotstyle=o](3,0)(3.25,0)
(3.5,0)(3.75,0)(4,0)(4.25,0)(4.5,0)(4.75,0)(5,0)(5.25,0)(5.5,0)(5.75,0)(6,0)(6.25,0)(6.5,0)
(6.75,0)(7,0)
\rput[c](3,-.28){$\mbox{\footnotesize$c_l\leq c_h/2$}$}
\rput[c](7,-.28){$\mbox{\footnotesize$c_h\geq (\overline{p}+c_l)/2$}$}
\psline{|<->|}(3,-.6)(7,-.6)
\psline(5,-.6)(5,-1)(5.2,-1)
\rput[l](5.35,-1){\footnotesize Nash range}
\psline{|<->|}(3,0.3)(4.6667,0.3)
\psline(3.875,0.3)(3.875,1.175)(3.675,1.735)(3.575,1.735)
\psline(3.875,1.175)(3.675,0.615)(3.575,.615)
\rput[r](3.5,1.7){\footnotesize Empirically plausible behavior}
\rput[r](3.5,1.35){\footnotesize in WB can approximate only }
\rput[r](3.5,1){\footnotesize equilibria to the left of }
\rput[r](3.5,.65){\footnotesize $\mbox{\footnotesize$2c_h/3$}$ in the Nash range}
\psline{|<->|}(5.3333,0.3)(7,0.3)
\psline(6.125,0.3)(6.125,1.175)(6.325,1.735)(6.425,1.735)
\psline(6.125,1.175)(6.325,0.615)(6.425,.615)
\rput[l](6.55,1.7){\footnotesize Empirically plausible behavior}
\rput[l](6.55,1.35){\footnotesize in LB can approximate only}
\rput[l](6.55,1){\footnotesize equilibria to the right of}
\rput[l](6.55,.65){\footnotesize $2c_l/3+\overline{p}/3$ in the Nash range}
\rput[c](5.5,-1.5){\footnotesize (a)}
\end{pspicture}
\begin{pspicture}(-1,-1.7)(11,2)
\psline{<->}(-.5,0)(10.5,0)
\psdots[dotsize=5pt,dotstyle=|](0,0)(10,0)
\rput[c](0,-.28){$\mbox{\footnotesize$0$}$}
\rput[c](10,-.28){$\mbox{\footnotesize$\overline{p}$}$}
\psdots[dotsize=4pt,dotstyle=o](3.5,0)(3.75,0)(4,0)(4.25,0)(4.5,0)(4.75,0)(5,0)(5.25,0)(5.5,0)(5.75,0)(6,0)(6.25,0)(6.5,0)
\rput[c](3.5,-.28){$\mbox{\footnotesize$c_l\geq c_h/2$}$}
\rput[c](6.5,-.28){$\mbox{\footnotesize$c_h\leq (\overline{p}+c_l)/2$}$}
\psline{|<->|}(3.5,-.6)(6.5,-.6)
\psline(5,-.6)(5,-1)(5.2,-1)
\rput[l](5.35,-1){\footnotesize Nash range}
\psline{|<->|}(3.5,0.3)(4.5,0.3)
\psline(3.875,0.3)(3.875,1)(3.675,1.55)(3.575,1.55)
\psline(3.875,1)(3.675,0.45)(3.575,.45)
\rput[r](3.5,1.35){\footnotesize Empirically plausible behavior}
\rput[r](3.5,1){\footnotesize in WB can approximate only}
\rput[r](3.5,.65){\footnotesize the left third of Nash range}
\psline{|<->|}(5.5,0.3)(6.5,0.3)
\psline(6.125,0.3)(6.125,1)(6.325,1.55)(6.425,1.55)
\psline(6.125,1)(6.325,0.45)(6.425,.45)
\rput[l](6.55,1.35){\footnotesize Empirically plausible behavior}
\rput[l](6.55,1){\footnotesize in LB can approximate only}
\rput[l](6.55,.65){\footnotesize the right third of Nash range}
\rput[c](5.5,-1.5){\footnotesize (b)}
\end{pspicture}
\caption{\textbf{Bias of extreme-price auctions.} Each Nash equilibrium of an extreme-price auction can be characterized by its payoff-determinant bid in the Nash range, i.e., the set $\{c_l,...,c_h\}$ (Proposition~\ref{Prop:charc-equili-auctions}).  Agent $l$ prefers equilibria to the right of the Nash range and agent $h$ prefers equilibria to the left of the Nash range. Empirically plausible behavior in the extreme price auctions, i.e., behavior that can be fit by an rQRE, cannot approximate all Nash equilibria. In WB only a left segment of the Nash range can be approximated (Theorem~\ref{Thm:characterization-ee-WB}). In LB only a right segment of the Nash range can be approximated (Theorem~\ref{Thm:characterization-ee-LB}). The length of these segments depends on the number of bids that are available to the left and right of the Nash range, respectively. Two cases can be distinguished for each auction. For WB it depends whether $c_l\leq c_h/2$. For LB it depends whether $c_h\geq (c_l+\overline{p})/2$.}\label{Fig:bias-auctions}
\end{figure}
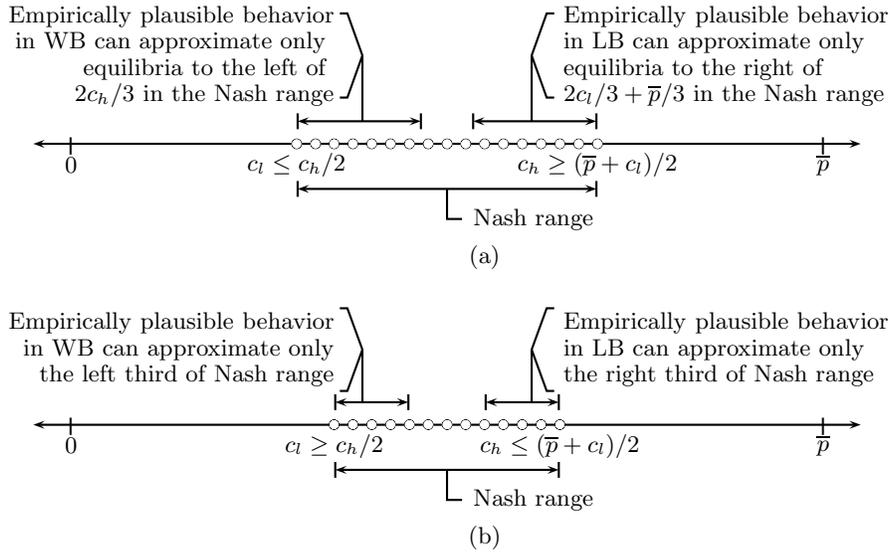

Theorems~\ref{Thm:characterization-ee-WB} and~\ref{Thm:characterization-ee-LB} allow us to conclude that we should expect the extreme-price auctions to be biased in opposite forms when they are operated. Observe that
\[\max\{2c_h/3-c_l,ES(v)/3\}=\left\{\begin{array}{ll}2c_h/3-c_l&\textrm{if }v_l\leq v_h/2;\\
ES(v)/3&\textrm{if }v_l\geq v_h/2.\end{array}\right.\]
When there is few bids to the left of the Nash range (i.e., $v_l\leq v_h/2$), empirically plausible behavior in WB can approximate only equilibria to the left of $2c_h/3$ in the Nash range.  If there is enough bids to the left of the Nash range (i.e., $v_l\geq v_h/2$), the bias of WB becomes much more pronounced: Empirically plausible behavior in WB that approximates a Nash equilibrium eventually assigns the HV agent essentially at least two thirds of the equity surplus. Symmetric statements hold for LB. (Figure~\ref{Fig:bias-auctions}).

Summarizing, if we suppose that behavior will be weakly payoff monotone when WB and LB are operated, as empirical distributions move towards mutual best responses,\footnote{One can think for instance, but not necessarily, of agents exhibiting less noisy best responses as they understand better the strategic situations.} these auctions will eventually exhibit opposed forms of bias: WB will benefit the HV agent and LB will benefit the LV agent.

Empirical equilibrium analysis predicts no bias for interior-price auctions in the following sense: For each payoff in the Nash range, there is a sequence of empirically plausible behavior that converges to an equilibrium with this payoff.

\begin{remark}\label{Prop:interior-price-Nequ}Consider a valuation structure $v$ and $p\in\{c_l,...,c_h\}$. For each interior price auction with valuations $v$, there is a Nash equilibrium, $\sigma$, such that $\pi_l(\sigma)=c_l+(p-c_l)$ and $\pi_h(\sigma)=c_h+(c_h-p)$, and a sequence of interior weakly payoff monotone behavior of $v$ that converges to $\sigma$.
\end{remark}

Proposition~\ref{Prop:strict-equil} states that each bid in the Nash range is the outcome of a strict Nash equilibrium.  Thus, to prove this remark, we only need to observe that a strict equilibrium is itself a weak payoff monotone distribution. Thus, it is trivially approached a constant sequence of weakly payoff monotone behavior. Now, the limit of a sequence of weakly payoff monotone behavior is always the limit of a sequence interior weakly payoff monotone distributions \citep{Velez-Brown-2018-EE}.

Remark~\ref{Prop:interior-price-Nequ} vis a vis Theorems~\ref{Thm:characterization-ee-WB} and~\ref{Thm:characterization-ee-LB} provides a meaningful insight about the performance of $\alpha$-auctions: While empirical plausibility of behavior and better understanding of the auction game forces WB and LB to be biased, no such conclusion can be made with any interior-price auction. It is worth noting that Proposition~\ref{Prop:interior-price-Nequ}  does not predict that interior-price auctions will always achieve more balanced divisions of the equity surplus than the extreme price auctions, however.

Finally, it is also interesting to determine the implications of empirical plausibility of equilibria for the efficiency of $\alpha$-auctions. We know that all Nash equilibria of extreme-price auctions are efficient (Proposition~\ref{Prop:charc-equili-auctions}).\footnote{This conclusion is essentially unchanged for the symmetric tie-breaker \citep{Velez-Brown-2019-EBA}.} Thus, one can expect that when extreme price auctions are operated, as empirical distributions of behavior move toward best responses, the auctions will approximate an efficient allocation. This is also possible for each interior price auction (Proposition~\ref{Prop:strict-equil}).

\begin{figure}[t]
\center
\savedata{\LBone}[(0,50.3)
(0.01,53.4)
(0.02,56.3)
(0.03,58.8)
(0.04,61.1)
(0.05,63.1)
(0.06,64.9)
(0.07,66.5)
(0.08,68)
(0.09,69.4)
(0.1,70.6)
(0.11,71.8)
(0.12,72.8)
(0.13,73.8)
(0.14,74.7)
(0.15,75.5)
(0.16,76.3)
(0.17,77)
(0.18,77.7)
(0.19,78.4)
(0.2,79)
(0.21,79.6)
(0.22,80.1)
(0.23,80.6)
(0.24,81.1)
(0.25,81.6)
(0.26,82)
(0.27,82.4)
(0.28,82.8)
(0.29,83.2)
(0.3,83.6)
]
\savedata{\ABone}[(0,50.3)
(0.01,53.5)
(0.02,56.6)
(0.03,59.6)
(0.04,62.4)
(0.05,65.2)
(0.06,67.8)
(0.07,70.3)
(0.08,72.6)
(0.09,74.7)
(0.1,76.6)
(0.11,78.4)
(0.12,80.1)
(0.13,81.6)
(0.14,83)
(0.15,84.2)
(0.16,85.4)
(0.17,86.4)
(0.18,87.4)
(0.19,88.3)
(0.2,89.1)
(0.21,89.8)
(0.22,90.5)
(0.23,91.1)
(0.24,91.7)
(0.25,92.2)
(0.26,92.7)
(0.27,93.1)
(0.28,93.6)
(0.29,93.9)
(0.3,94.3)
]
\savedata{\WBone}[
(0,50.3)
(0.01,53.6)
(0.02,56.8)
(0.03,60)
(0.04,63)
(0.05,65.9)
(0.06,68.6)
(0.07,71)
(0.08,73.3)
(0.09,75.3)
(0.1,77.2)
(0.11,78.8)
(0.12,80.3)
(0.13,81.7)
(0.14,82.9)
(0.15,84)
(0.16,85)
(0.17,85.9)
(0.18,86.7)
(0.19,87.4)
(0.2,88.1)
(0.21,88.7)
(0.22,89.3)
(0.23,89.8)
(0.24,90.3)
(0.25,90.7)
(0.26,91.1)
(0.27,91.5)
(0.28,91.8)
(0.29,92.1)
(0.3,92.4)
]
\psset{lly=-0.5cm, xAxisLabel={$\lambda$},yAxisLabel={\% Efficient allocations}, xAxisLabelPos={c, -0.3in},yAxisLabelPos={-0.45in,c},ysubticks=2,
    xsubticks=10,subticksize=0.5}
\psset{showpoints=true, linewidth=0.8pt,dotscale=1}
\begin{pspicture}(0,0)(1,5)
\end{pspicture}
\begin{psgraph}[axesstyle=frame,Ox=0, Oy=0, Dy=25, Dx=0.1](0,0)(0.301,100){4.5cm}{4cm}
\listplot[linestyle=dashed,plotstyle=curve,showpoints=false]\WBone
\listplot[linecolor=gray,plotstyle=curve,showpoints=false]\ABone
\listplot[linestyle=dotted,dotsep=1pt,plotstyle=curve,showpoints=false]\LBone
\psline[linestyle=dotted,showpoints=false](0,25)(.3,25)
\psline[linestyle=dotted,showpoints=false](0,50)(.3,50)
\psline[linestyle=dotted,showpoints=false](0,75)(.3,75)
\psline[showpoints=false,linewidth=0.8pt](0,100)(0,116.67)(0.301,116.67)(.301,100)
\rput(0.03,108.33){\psline[linestyle=dashed,showpoints=false](0,0)(0.5cm,0)}
\rput[l](.0675,108.33){\footnotesize{WB}}
\rput(.12,108.33){\psline[linecolor=gray,showpoints=false](0,0)(0.5cm,0)}
\rput[l](.1575,108.33){\footnotesize{AB}}
\rput(.2025,108.33){\psline[dotsep=1pt,showpoints=false](0,0)(0.5cm,0)}
\rput[l](.24,108.33){\footnotesize{LB}}
\rput[c](.15,125){Valuation structure 1A}
\end{psgraph}
\savedata{\LBtwo}[
(0,50.2)
(0.01,53.3)
(0.02,56.1)
(0.03,58.5)
(0.04,60.6)
(0.05,62.5)
(0.06,64.2)
(0.07,65.8)
(0.08,67.2)
(0.09,68.5)
(0.1,69.7)
(0.11,70.8)
(0.12,71.9)
(0.13,72.8)
(0.14,73.7)
(0.15,74.6)
(0.16,75.3)
(0.17,76.1)
(0.18,76.8)
(0.19,77.4)
(0.2,78)
(0.21,78.6)
(0.22,79.2)
(0.23,79.7)
(0.24,80.2)
(0.25,80.6)
(0.26,81.1)
(0.27,81.5)
(0.28,81.9)
(0.29,82.3)
(0.3,82.7)
]
\savedata{\ABtwo}[
(0,50.2)
(0.01,53.3)
(0.02,56.4)
(0.03,59.3)
(0.04,62.2)
(0.05,64.9)
(0.06,67.5)
(0.07,69.9)
(0.08,72.2)
(0.09,74.3)
(0.1,76.2)
(0.11,78.1)
(0.12,79.7)
(0.13,81.2)
(0.14,82.6)
(0.15,83.9)
(0.16,85.1)
(0.17,86.2)
(0.18,87.2)
(0.19,88.1)
(0.2,88.9)
(0.21,89.7)
(0.22,90.3)
(0.23,91)
(0.24,91.6)
(0.25,92.1)
(0.26,92.6)
(0.27,93.1)
(0.28,93.5)
(0.29,93.9)
(0.3,94.2)
]
\savedata{\WBtwo}[
(0,50.2)
(0.01,53.4)
(0.02,56.4)
(0.03,59.1)
(0.04,61.5)
(0.05,63.7)
(0.06,65.6)
(0.07,67.3)
(0.08,68.9)
(0.09,70.3)
(0.1,71.6)
(0.11,72.7)
(0.12,73.8)
(0.13,74.8)
(0.14,75.7)
(0.15,76.6)
(0.16,77.3)
(0.17,78.1)
(0.18,78.8)
(0.19,79.4)
(0.2,80)
(0.21,80.6)
(0.22,81.1)
(0.23,81.6)
(0.24,82.1)
(0.25,82.5)
(0.26,83)
(0.27,83.4)
(0.28,83.8)
(0.29,84.1)
(0.3,84.5)
]
\begin{pspicture}(0,0)(2,5)
\end{pspicture}
\begin{psgraph}[axesstyle=frame,Ox=0, Oy=0, Dy=25, Dx=0.1](0,0)(0.301,100){4.5cm}{4cm}
\listplot[linestyle=dashed,plotstyle=curve,showpoints=false]\WBtwo
\listplot[linecolor=gray,plotstyle=curve,showpoints=false]\ABtwo
\listplot[plotstyle=curve,showpoints=false]\LBtwo
\psline[linestyle=dotted,showpoints=false](0,25)(.3,25)
\psline[linestyle=dotted,showpoints=false](0,50)(.3,50)
\psline[linestyle=dotted,showpoints=false](0,75)(.3,75)
\psline[showpoints=false,linewidth=0.8pt](0,100)(0,116.67)(0.301,116.67)(.301,100)
\rput(0.03,108.33){\psline[linestyle=dashed,showpoints=false](0,0)(0.5cm,0)}
\rput[l](.0675,108.33){\footnotesize{WB}}
\rput(.12,108.33){\psline[linecolor=gray,showpoints=false](0,0)(0.5cm,0)}
\rput[l](.1575,108.33){\footnotesize{AB}}
\rput(.2025,108.33){\psline[dotsep=1pt,showpoints=false](0,0)(0.5cm,0)}
\rput[l](.24,108.33){\footnotesize{LB}}
\rput[c](.15,125){Valuation structure 2A}
\end{psgraph}
\vspace{.5cm}
\caption{\% of efficient allocations in a symmetric Logistic QRE for WB, AB, and LB auctions \citep[see][for definitions]{Goeree-Holt-Palfrey-2016-Book}. Valuations and bid range correspond to two environments implemented in our experiments. In valuation structure 1A, $\Bcal=\{0,..,160\}$, $c_l=10$, and $c_h=30$. In valuation structure 2A, $\Bcal=\{0,..,290\}$, $c_l=100$, and $c_h=120$.}\label{Fig:LQRE-calculations1}
\end{figure}
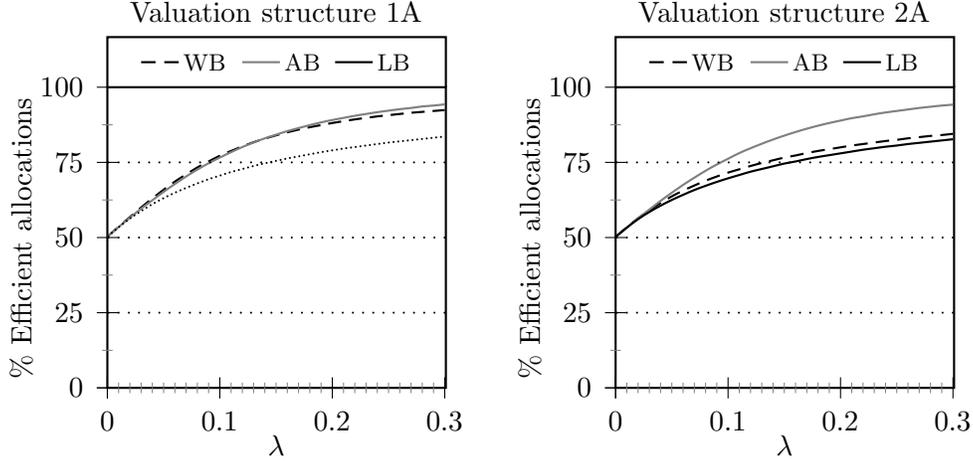

A quantitative exercise allows us to visualize a pattern in which this process may occur. Fig.~\ref{Fig:LQRE-calculations1} shows the percentage of efficient allocations in a symmetric Logistic QRE in the WB, AB, and LB auctions, for two valuation structures. These models satisfy weak payoff monotonicity.  In each environment efficiency increases as $\lambda$, the model's proxy for agents' understanding of the mechanism, increases. Interestingly, which auction maximizes efficiency for a given $\lambda$ depends on the structural parameters of the game, i.e., the position of the Nash range within the set of possible bids.

One can conclude that it is an empirical question to determine what auction format performs best in a given environment. It is reasonable to hypothesize that the mechanism that is understood the best is the one that achieves higher efficiency.

\section{Experimental design and procedures}\label{Sec:Exp-design}

We design an experiment to test the comparative statics predicted by Theorems~\ref{Thm:characterization-ee-WB} and~\ref{Thm:characterization-ee-LB} and to evaluate the general performance of $\alpha$-auctions. %
%
We test three different mechanisms: WB, AB, and LB. We test each mechanism under several different valuation structures which vary by session. Importantly, each type of session-valuation is balanced equally across all three treatments. So there are equal and equivalent numbers of sessions in each of the three treatments. 

\subsection{Experimental design}
The experiment implemented the environment and $\alpha$-auction described in Sec.~\ref{Sec:model}. The $3\times 1$ design utilized parameter values of $\alpha=0, 0.5, 1$, to produce WB, AP, and LB respectively. Subjects, randomly selected into groups of two, determined how to allocate two indivisible items with possible transfer payments.\footnote{Consistent with previous literature \cite[i.e.,][]{Brown-Velez-2016-GEB}, we chose to have subjects bid over two items to match the generalized theoretical environment more closely and reduce the possibility that subjects are motivated by the non-monetary desire to ``win'' an item \citep[e.g.,][]{cooper2008understanding,Roider-2012-SJE}.} In all possible allocations, each subject received \emph{exactly} one item. Subjects received points for acquiring an item, equal to their value of that item (i.e., induced values) plus or minus any points they transferred to the other subject. Subjects' values of both items were common knowledge to both players.

Under the $\alpha$-auction, subjects submitted their bids for the item. The subject with the higher bid received the item, and the subject with the lower bid received a transfer equal to a convex combination of both bids, determined by the $\alpha$ value associated with the auction. That is, the transfer amount equaled $\alpha\times[\text{high bid}]+(1-\alpha)\times[\text{low bid}]$. 
In the case of equal bids, the item was assigned to the subject with the HV for item B.

\begin{figure}
\centering
\includegraphics[width=1\textwidth]{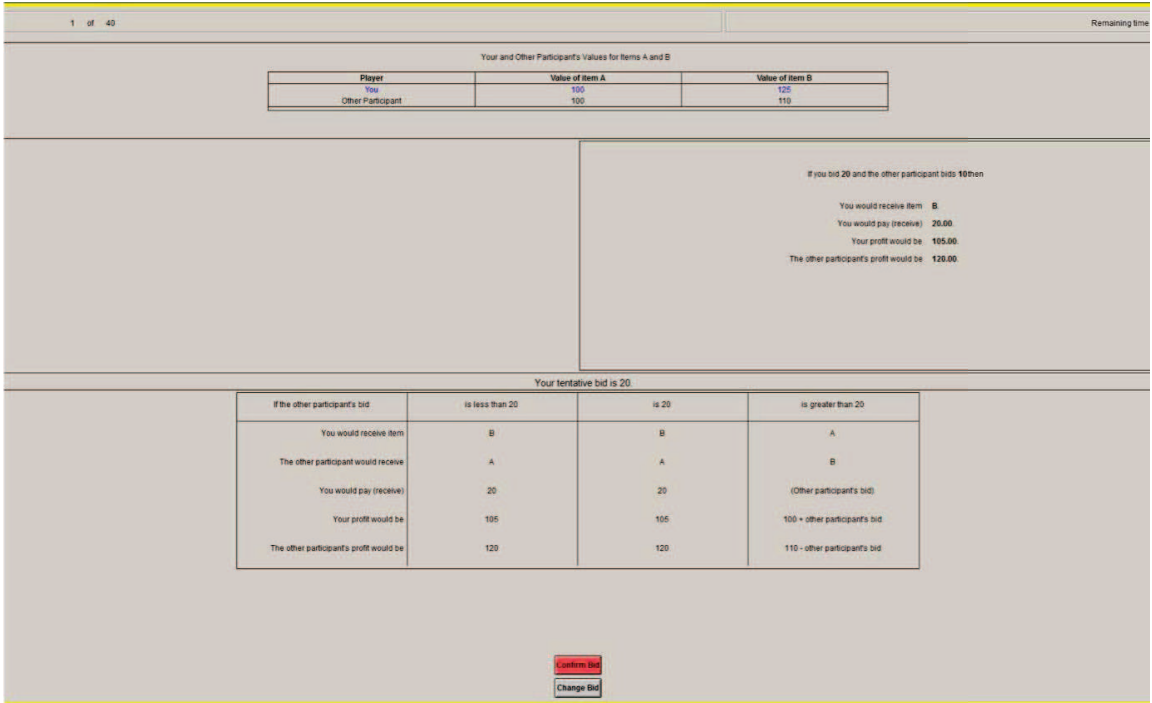}
\caption{The $\alpha$-auction Interface. Winner's Bid Auction ($\alpha=1$) treatment shown.}
\label{WBA}
\end{figure}

After submitting a bid, each subject was allowed to submit a possible value for the other player's bid. The experimental software then displayed the outcome (i.e., who gets which item, what amount is transferred for each player, each players' earnings for that period) that would occur with those two bids as well as a table that showed all possibilities that could happen if the other player's bid were below, equal to, or above the subject's bid (see Figure \ref{WBA}). After a subject viewed these possibilities, she could choose to confirm her bid, or chose an alternate bid. If she chose an alternate bid, the process repeated. The process ended when a subject confirmed her bid.

At the end of the game a feedback screen would describe both players actions under the mechanism and provide information on each subject's valuation and total points earned that round. All information was revealed to subjects at the end of the game as feedback with the intent to aide learning over the course of the session. At the end of each game, subjects would be reassigned to a subject pair and a new break-up game would begin with each subject drawing new valuations. This process would continue for 40 periods.

Both players had a constant and equal value on item A that remained the same throughout the session. For any period, for each grouping of subjects, one subject was randomly assigned the high value on item $B$, the other was assigned the low value on item $B$.  Thus a subject's value on item~$B$ could change for any period. The pair of valuations on item $B$ generally remained constant. Either they changed once between period 20 and 21, halfway through the session, or they did not change at all.

To avoid incentives associated with repeated play, subjects were randomly re-assigned to each other at the beginning of each period.  Subjects were instructed that they were to be randomly rematched each period, and no identifying information (e.g., subject number) was disclosed to a subject about her match in any round.  Each period began with each subject seeing the valuation structure for the period.


\subsection{Experimental Procedures}

\begin{table}[t]
\centering
\begin{tabular}{ccc}\hline
Player & \begin{tabular}{cc}Value of\\ Item~$A$\end{tabular}& \begin{tabular}{cc}Value of\\ item~$B$\end{tabular}\\\hline
Player 1 & $50$ & $250$\\
Player 2 & $50$ & $290$\\\hline
\end{tabular}
\caption{\textbf{A sample valuation.} \label{sample valuation}In this valuation, one subject valued item~$A$ at 50 and item~$B$ at 250.  As in our theoretical environment
, the other subject with whom she is paired had the same value for item~$A$, but valued item~$B$ at 290.  Each player had an equal chance of receiving the high value on item~$B$ for any period.  Values were common knowledge to both players.} 
\end{table}

Twelve sessions were held at the Economic Research Laboratory (ERL) in the Economics Department at Texas A\&M University during spring and summer 2016. Subjects were recruited using ORSEE software \citep{Greiner_2015} and made their decisions on software programmed in the Z-tree language \citep{fischbacher}. Subjects sat at computer terminals with dividers to make sure their anonymity was preserved. Subjects were 246 Texas A\&M undergraduates from a variety of majors.

Experimental sessions can be categorized under four valuation structures, each featuring one session under the WB, LB and AP auctions. In valuation structure 1, item A was valued at 100 by both players; item B had respective valuations of $(120, 160)$ for the first twenty periods and $(120, 320)$ for the second twenty periods. In the remaining three structures all agents valued item A at 50. In valuation structure 2, item B had respective valuations of $(250, 290)$ for the first twenty periods and $(250, 450)$ for the second twenty periods. In valuation structure 3, item B had respective valuations of $(250, 450)$ for all forty periods. In valuation structure 4, item B had respective valuations of $(250, 290)$ for all forty periods. Sessions consisted of between 14--30 subjects. See Table \ref{procedures} for more detail.

Experiments lasted about two hours. To avoid issues with preferences that involve complementarities across periods \cite[see][]{azrieli_2014,Brown-Healy-2018-EER}, one period was randomly selected at the end of each experiment to be paid. Subjects received earnings from that round converted to cash at the rate of 1 point=\$0.10 (or 1 point=\$0.13 under valuation structure 1) plus a \$5 show-up payment. Earnings ranged from \$10.50 to \$44.00 with \$25.87 average.

%

\begin{table}\centering
\resizebox{1\textwidth}{!} {
\begin{tabular}{cccccccc}
\toprule
\begin{tabular}{c}session\\ type\end{tabular}&\begin{tabular}{c}valuation\\ structure\end{tabular}&periods& \begin{tabular}{c}value of\\item A\end{tabular}& \begin{tabular}{c}values of\\item B\end{tabular} &\begin{tabular}{c}WB\\subjects\end{tabular}&\begin{tabular}{c}AP\\subjects\end{tabular}&\begin{tabular}{c}LB\\subjects\end{tabular}\\
\midrule
\multirow{2}{*}{1}&1A&1--20&100&(120,160)&\multirow{2}{*}{20}&\multirow{2}{*}{30}&\multirow{2}{*}{22}\\
&1B&21--40&100&(120,320)&&&\\
\multirow{2}{*}{2}&2A&1--20&50&(250,290)&\multirow{2}{*}{20}&\multirow{2}{*}{20}&\multirow{2}{*}{20}\\
&2B&21--40&50&(250,450)&&&\\
3&3&1--40&50&(250,450)&20&20&20\\
4&4&1--40&50&(250,290)&20&14&18\\
\bottomrule
\end{tabular}
}
\caption{\label{procedures} Valuation structures used in 12 experimental sessions. Four different session types were used over the three auction mechanism treatments. The first two session types used two different valuation structures in Periods 1--20 and 21--40 respectively. Session types 3 and 4 used on valuation structure for the entire session. There are six total valuation structures. 246 total subjects participated in these 12 sessions.}
\end{table}

\section{Experimental results}\label{Sec:Exp-results}
\subsection{Mechanism bias}\label{Sec:mechanism-bias}
The quintessential implication of our theoretical model in this environment is that WB and LB---essentially equivalent in terms of Nash equilibrium predictions---have profoundly different implications for the HV and LV bidder. In short, WB will only achieve equilibria that are most favorable to the HV agent, and LB will only achieve equilibria that are most favorable to the LV agent. This is effectively mechanism bias.

To make results comparable across all sessions and valuations we focus on standardized payoffs. Specifically we calculate equity surpluses above maximin. Recall that any agent can ensure a minimum payoff by making a specific bid. The payoff and bid do not vary by auction mechanism. This is the maximin bid and the minimum payoff guaranteed is the maximin payoff. 
Because  HV and LV agents have different maximin bids, these values are different for each bidder type.
\begin{equation}
\tilde{\pi}_i=\left\{\begin{array}{ll} \frac{\pi_i-v_h/2}{v_h/2-v_l/2}=\frac{\pi_i-v_h/2}{ES(v)} & \text{if } v_i=v_h,\\
\frac{\pi_i-v_l/2}{v_h/2-v_l/2}=\frac{\pi_i-v_l/2}{ES(v)} & \text{if } v_i=v_l.
\end{array}
\right.
\end{equation}

A general property of these standardized payoffs is that when an outcome is efficient the HV and LV agents' values will total to 1. That is, the equity surplus is fully shared between both agents. If the outcome is inefficient, the value of payoffs will total to -1. Thus totaling standardized payoff values and comparing to 1 reveals the proportion of efficient outcomes achieved under the mechanism. (Totals above 0 indicate more than half the outcomes were efficient).

As we will see in the results, generally if equity divisions are very favorable to one type of agent in efficient outcomes, they will be unfavorable to the same agent in inefficient outcomes. To see why, consider a HV agent that gets 95\% of the equity surplus when outcomes are efficient. This means the item is trading at only 5\% over the LV maximin. Should an inefficient outcome occurs, the HV agent receives no item (a payoff of 0) and receives only 5\% of the equity surplus as compensation. This is 0.95 equity surpluses worse than their maximin.

Table \ref{tab:payoffs} provides overall summary statistics of standardized payoffs (equity surpluses above maximin payoff) by valuation-session level for our eight sessions (12 valuations total) of WB and LB. (Recall that in the first two session-trios, subjects experienced twenty periods of two consecutive valuations, hence the A and B). Whether we look at the valuation or session level, in all cases the party predicted by theory to have higher payoffs has higher payoffs. High-value bidders receive 0.225--0.664 surpluses above their maximin in WB compared to -0.171--0.099 for the LV bidder.  In contrast, HV bidders receive -0.022--0.221 equity surpluses above their maximin compared to 0.262--0.410 for the LV bidder in LB. Simple binary permutation tests---either considering the 8 or 12 possible binomial draws of the null hypothesis depending on whether we think about this at the session or valuation level---would find this result significant at the 1\%-level.
\begin{table}[t]
\centering
\resizebox{0.9\textwidth}{!} {
\begin{tabular}{lccccccc}
\toprule
&\multicolumn{3}{c}{HV bidder}&&\multicolumn{3}{c}{LV bidder}\\
\begin{tabular}{c}valuation\\ structure\end{tabular}& WB & AP &LB& &WB& AP & LB\\
\midrule
\begin{tabular}{c}1A\\(20 periods)\end{tabular}& \begin{tabular}{c}0.336\\(0.881)\\200\end{tabular}&\begin{tabular}{c}0.140\\(0.651)\\300\end{tabular}&\begin{tabular}{c}0.038\\(0.481)\\220\end{tabular}&&\begin{tabular}{c}0.105\\(0.933)\\300\end{tabular}&\begin{tabular}{c}0.247\\(0.835)\\220\end{tabular}&\begin{tabular}{c}0.262\\(0.847)\\300\end{tabular}\\
\vfivefive
\begin{tabular}{c}1B\\(20 periods)\end{tabular}& \begin{tabular}{c}0.664\\(0.465)\\200\end{tabular}&\begin{tabular}{c}0.312\\(0.483)\\300\end{tabular}&\begin{tabular}{c}0.099\\(0.268)\\220\end{tabular}&&\begin{tabular}{c}0.176\\(0.175)\\300\end{tabular}&\begin{tabular}{c}0.248\\(0.414)\\220\end{tabular}&\begin{tabular}{c}0.410\\(0.763)\\300\end{tabular}\\
\vfivefive
\begin{tabular}{c}2A\\(20 periods)\end{tabular}& \begin{tabular}{c}0.225\\(1.665)\\260\end{tabular}&\begin{tabular}{c}0.217\\(1.384)\\180\end{tabular}&\begin{tabular}{c}-0.112\\(2.277)\\180\end{tabular}&&\begin{tabular}{c}0.091\\(1.321)\\260\end{tabular}&\begin{tabular}{c}-0.006\\(1.043)\\180\end{tabular}&\begin{tabular}{c}0.257\\(2.259)\\180\end{tabular}\\
\vfivefive
\begin{tabular}{c}2B\\(20 periods)\end{tabular}& \begin{tabular}{c}0.441\\(0.553)\\260\end{tabular}&\begin{tabular}{c}0.215\\(0.335)\\180\end{tabular}&\begin{tabular}{c}0.023\\(0.464)\\180\end{tabular}&&\begin{tabular}{c}0.221\\(0.311)\\260\end{tabular}&\begin{tabular}{c}0.429\\(0.513)\\180\end{tabular}&\begin{tabular}{c}0.355\\(0.793)\\180\end{tabular}\\
\vfivefive
\begin{tabular}{c}3\\(40 periods)\end{tabular}& \begin{tabular}{c}0.432\\(0.752)\\400\end{tabular}&\begin{tabular}{c}0.215\\(0.574)\\400\end{tabular}&\begin{tabular}{c}0.022\\(0.478)\\400\end{tabular}&&\begin{tabular}{c}0.203\\(0.361)\\400\end{tabular}&\begin{tabular}{c}0.330\\(0.549)\\400\end{tabular}&\begin{tabular}{c}0.348\\(0.896)\\400\end{tabular}\\
\vfivefive
\begin{tabular}{c}4\\(40 periods)\end{tabular}& \begin{tabular}{c}0.381\\(1.440)\\400\end{tabular}&\begin{tabular}{c}0.187\\(2.065)\\280\end{tabular}&\begin{tabular}{c}-0.171\\(1.372)\\360\end{tabular}&&\begin{tabular}{c}-0.022\\(0.933)\\400\end{tabular}&\begin{tabular}{c}0.013\\(1.718)\\280\end{tabular}&\begin{tabular}{c}0.349\\(1.675)\\360\end{tabular}\\
\midrule
\begin{tabular}{c}overall\end{tabular}& \begin{tabular}{c}0.406\\(1.098)\\1,720\end{tabular}&\begin{tabular}{c}0.215\\(1.072)\\1,640\end{tabular}&\begin{tabular}{c}-0.025\\(1.079)\\1,560\end{tabular}&&\begin{tabular}{c}0.122\\(0.789)\\1,720\end{tabular}&\begin{tabular}{c}0.220\\(0.950)\\1,640\end{tabular}&\begin{tabular}{c}0.335\\(1.302)\\1,560\end{tabular}\\
\bottomrule
\end{tabular}
}
\caption{\label{tab:payoffs} Mean, standard deviation, and number of observations of standardized payoffs, by auction type, session$\times$valuation structure for HV and LV bidders.}
\end{table}
Of course, observing aggregate results at the session level may be problematic if there is substantial within-session variation that is ignored \cite[see][]{frechette}. For this reason, we also consider mechanism bias at the subject level. Because our experiment randomized whether a subject was HV or LV each period, we can provide within-subject evidence for mechanism bias. Non-parametric results confirm our findings. Of the 164 subjects in either WB or LB, 134 (82\%) achieved higher payoffs as the higher (or lower) valuation bidder in the direction predicted by theory. A binomial test shows the result is significant at the 1\% level.

We use regression analysis to parametrically model payoffs across mechanisms. Table~\ref{tab:payreg} provides regression analysis of standardized payoffs across auction mechanisms for all outcomes, and only efficient and inefficient outcomes. The overall results are similar to what can be observed from session averages. In WB the HV bidder receives payoffs of 0.406 equity surpluses above her maximin (the sum of coefficients WB auction + HV bidder and their interaction term), the LV agent receives 0.123 equity surpluses above her maximin (a difference of 0.284, statistically different than 0, $p<0.01$). In LB this relation is reversed. As predicted by theory, the LV bidder receives 0.335 surpluses above maximin statistically different than the HV bidder who receives -0.025 ($p<0.01$). In AB the LV bidder receives 0.220 (the constant) and the HV bidder receives a roughly identical amount.

\begin{table}[t]
\centering
\begin{tabular}{lccc} \toprule
 & (1) & (2) & (3) \\
&\multicolumn{3}{c}{\begin{tabular}{c}standardized earnings\\(equity surpluses over maximin)\end{tabular}}\\
& all outcomes & efficient & not efficient\\ \midrule
 &  &  &  \\
WB  & -0.098** & -0.249*** & 0.253** \\
& (0.038) & (0.050) & (0.103) \\
LB& 0.115** & 0.466*** & -0.426*** \\
& (0.051) & (0.067) & (0.089) \\
HV bidder & -0.005 & 0.168** & -0.442*** \\
 & (0.040) & (0.065) & (0.123) \\
WB $\times$& 0.289*** & 0.497*** & -0.506*** \\
HV bidder& (0.054) & (0.081) & (0.160) \\
LB$\times$& -0.354*** & -0.932*** & 0.852*** \\
HV bidder& (0.058) & (0.101) & (0.152) \\
Constant & 0.220*** & 0.416*** & -0.279*** \\
 & (0.030) & (0.043) & (0.064) \\
 &  &  &  \\
\midrule
Observations & 9,840 & 7,024 & 2,816 \\
 R-squared & 0.017 & 0.110 & 0.075 \\ \bottomrule
\multicolumn{4}{c}{ *** p$<$0.01, ** p$<$0.05, * p$<$0.1} \\
\end{tabular}
\caption{Regressions of standardized payoffs on auction and bidder type for all allocations, and efficient/inefficient allocations. Standard errors are clustered at the subject level.\label{tab:payreg}}
\end{table}

Restricting our attention to efficient outcomes is quite useful to discover the nature of the mechanism bias. Over time we expect efficient outcomes to become more prevalent; ultimately overall payoffs will converge to efficient payoffs if there is enough time and frequencies approach mutual best responses (Proposition~\ref{Prop:charc-equili-auctions}). The results emphasize our main findings. In efficient outcomes in the WB, the HV agent receives $2/3$ more of the equity surplus than the LV agent. In LB the LV agent receives 0.764 more of the equity surplus than the HV agent. Under AB  the HV agent receives 0.168 more. All results are statistically different across the auctions ($p<0.01$).

As expected, inefficient outcomes flip the advantage for the favored party. The HV agent receives -0.945 less than the LV agent in  WB. The LV agent receives 0.409 less than the HV agent in LB. In AB the HV bidder receives 0.442 less than the LV bidder.

\begin{figure}[t]\centering
\includegraphics[width=0.9\textwidth]{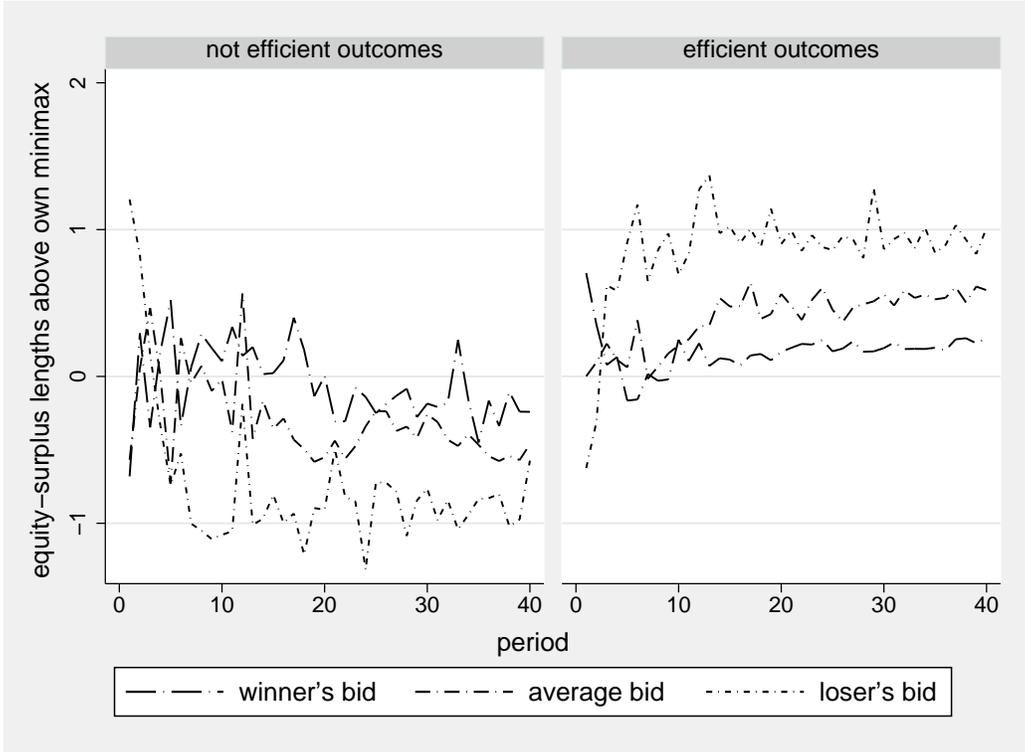}
\caption{\label{paytime2}Average normalized payoffs per period within the experiments for LV agent in efficient outcomes (left) and inefficient outcomes (right). In an efficient allocation the normalized payoffs add up to one. Thus, the graph of the normalized payoff for the HV agent in an efficient allocation is the mirror image, of the LV agent's, with respect to the horizontal line of level 0.5. In an efficient allocation the normalized payoffs add up to minus one. Thus, the graph of the normalized payoff for the HV agent in an inefficient allocation is the mirror image, of the LV agent's, with respect to the horizontal line of level -0.5.}
\end{figure}

Figure~\ref{paytime2} shows the time trends of standardized payoffs over the 40 periods of the experiment. The LV agent receives higher payoffs in LB and lowest payoffs in WB. These rankings are reversed for the HV agent. Payoffs for both agents in AB are in the middle of the other two auctions. These effects persist, if not intensify over the course of each session. Separating efficient and inefficient outcomes provides an even sharper separation of payoffs across mechanisms for both agents. Conditional on an efficient allocation, the ranking based on overall payoffs coincides with that based on overall payoffs (Figure~\ref{paytime2} (left)). Conditional on an inefficient allocation, the ranking is reversed (Figure~\ref{paytime2} (right)). The efficiency loss is mostly paid by the agent who is favored by each respective mechanism.

In short, we find substantial evidence for the mechanism bias in the direction consistent with theory. The sole mitigator of this bias is the achievement of inefficient outcomes which if bids are consistent would flip the bias. Because these inefficient outcomes are occur less often than inefficient ones and decrease over time, they are not sufficient to negate the implications of our main theoretical prediction.

\subsection{Bids}\label{Sec:bids}

Having identified the mechanism bias as predicted by Theorems~\ref{Thm:characterization-ee-WB} and~\ref{Thm:characterization-ee-LB}, we now proceed to unravel its causes. We first examine the direct cause of the bias that is also predicted by these theorems: the differences in bid distributions between mechanisms. WB should have the lowest bids and the LV auction the highest.

To make bids comparable across different valuation structures, we standardize them as well. Here we do not need to standardize HV and LV agents  bids differently. We define them as equity surpluses above the LV maximin. Conveniently, the Nash range for deterministic bids is $[0,1]$.
\begin{equation}
\tilde{b}_i=\frac{b_i-v_l/2}{v_h/2-v_l/2}=\frac{b_i-v_l/2}{ES(v)}
\end{equation}

Table \ref{tab:bids} provides valuation-level summary statistics for standardized bids and payoffs for HV and LV bidders, respectively. 
The results confirm the comparative static predictions of theory. In all six valuations, average bids are lowest in WB, next highest in AB, and highest in LB. If we think of each of these valuation-pairs as independent observations, we note that this is the most extreme of 46,656 ($6^6$) possible permutations supporting this hypothesis; a permutation test finds this result statistically significant $(p<0.01)$.\footnote{Alternatively if we think of each session as independent, this is the most extreme of 1296 ($6^4$) observations and still falls under the same level of significance ($p<0.01$).}
\begin{table}[t]
\centering
\resizebox{0.9\textwidth}{!} {
\begin{tabular}{lccccccc}
\toprule
&\multicolumn{3}{c}{HV bidder}&&\multicolumn{3}{c}{LV bidder}\\
\begin{tabular}{c}valuation\\structure\end{tabular}& WB & AP &LB& &WB& AP & LB\\
\midrule
\begin{tabular}{c}1A\\(20 periods)\end{tabular}& \begin{tabular}{c}0.309\\(0.568)\\200\end{tabular}&\begin{tabular}{c}0.760\\(0.797)\\300\end{tabular}&\begin{tabular}{c}1.482\\(1.312)\\220\end{tabular}&&\begin{tabular}{c}0.120\\(0.743)\\200\\\end{tabular}&\begin{tabular}{c}0.553\\(0.748)\\300\end{tabular}&\begin{tabular}{c}1.165\\(1.031)\\220\end{tabular}\\
\vfivefive
\begin{tabular}{c}1B\\(20 periods)\end{tabular}& \begin{tabular}{c}0.198\\(0.145)\\200\end{tabular}&\begin{tabular}{c}0.530\\(0.258)\\300\end{tabular}&\begin{tabular}{c}1.198\\(0.500)\\220\end{tabular}&&\begin{tabular}{c}0.018\\(0.094)\\200\end{tabular}&\begin{tabular}{c}0.372\\(0.222)\\300\end{tabular}&\begin{tabular}{c}0.900\\(0.294)\\220\end{tabular}\\
\vfivefive
\begin{tabular}{c}2A\\(20 periods)\end{tabular}& \begin{tabular}{c}-0.520\\(1.453)\\260\end{tabular}&\begin{tabular}{c}0.152\\(1.336)\\180\end{tabular}&\begin{tabular}{c}2.020\\(3.091)\\180\end{tabular}&&\begin{tabular}{c}-0.888\\(1.515)\\260\end{tabular}&\begin{tabular}{c}-0.023\\(1.397)\\180\end{tabular}&\begin{tabular}{c}1.766\\(2.921)\\180\end{tabular}\\
\vfivefive
\begin{tabular}{c}2B\\(20 periods)\end{tabular}& \begin{tabular}{c}0.292\\(0.195)\\260\end{tabular}&\begin{tabular}{c}0.787\\(0.276)\\180\end{tabular}&\begin{tabular}{c}1.316\\(0.867)\\180\end{tabular}&&\begin{tabular}{c}-0.034\\(0.383)\\260\end{tabular}&\begin{tabular}{c}0.501\\(0.229)\\180\end{tabular}&\begin{tabular}{c}0.943\\(0.550)\\180\end{tabular}\\
\vfivefive
\begin{tabular}{c}3\\(40 periods)\end{tabular}& \begin{tabular}{c}0.154\\(0.414)\\400\end{tabular}&\begin{tabular}{c}0.671\\(0.598)\\400\end{tabular}&\begin{tabular}{c}1.547\\(0.995)\\400\end{tabular}&&\begin{tabular}{c}-0.242\\(0.387)\\400\end{tabular}&\begin{tabular}{c}0.365\\(0.445)\\400\end{tabular}&\begin{tabular}{c}1.028\\(0.575)\\400\end{tabular}\\
\vfivefive
\begin{tabular}{c}4\\(40 periods)\end{tabular}& \begin{tabular}{c}-0.544\\(1.198)\\400\end{tabular}&\begin{tabular}{c}0.039\\(2.210)\\280\end{tabular}&\begin{tabular}{c}2.241\\(2.445)\\360\end{tabular}&&\begin{tabular}{c}-0.867\\(1.253)\\400\end{tabular}&\begin{tabular}{c}-0.382\\(2.133)\\280\end{tabular}&\begin{tabular}{c}2.022\\(2.192)\\360\end{tabular}\\
\midrule
\begin{tabular}{c}overall\end{tabular}& \begin{tabular}{c}-0.066\\(0.936)\\1,720\end{tabular}&\begin{tabular}{c}0.509\\(1.153)\\1,640\end{tabular}&\begin{tabular}{c}1.676\\(1.798)\\1,560\end{tabular}&&\begin{tabular}{c}-0.381\\(0.998)\\1,720\end{tabular}&\begin{tabular}{c}0.245\\(1.122)\\1,640\end{tabular}&\begin{tabular}{c}1.334\\(1.603)\\1,560\end{tabular}\\
\bottomrule
\end{tabular}
}
\caption{\label{tab:bids} Mean, standard deviation, and number of observations of standardized bids, by auction type, session$\times$valuation structure for HV and LV agents.}
\end{table}

Again, this session-level analysis may be problematic if there is substantial within-session variation that is ignored \cite[see][]{frechette}. We next examine bidding distributions at the subject level. Unlike our previous analysis of mechanism bias, we are primarily concerned with between-subject variation in different treatments, rather than within-subject differences for HV and LV agents. Table~\ref{bidregs} provides regression analysis. The results confirm our suspicion. Bidders in WB bid 0.576--0.627 equity surpluses lower than those in AB who bid 1.010--1.089 equity surpluses lower than those in LB ($p<0.01$ for both comparisons). These findings are consistent with the averages we note in Table~\ref{tab:bids}.

Also of note is the comparative statics involving the variance of the bidding distributions. Nash equilibria of extreme price auctions always involve an agent placing a deterministic bid: the HV agent in WB and the LV agent in LB (Proposition~\ref{Prop:charc-equili-auctions}). If weakly payoff-monotone behavior approximates one of these Nash equilibria, unless the equilibrium produces the extreme outcome of the Nash range, the other agent uses a strictly mixed strategy (Theorems~\ref{Thm:characterization-ee-WB} and~\ref{Thm:characterization-ee-LB}). Thus, empirical analysis predicts that bidding variance should be greater for the LV bidder in LB and variance should be greater for the HV bidder in WB.\footnote{Our results do not make predictions for the AP auction, but interestingly, the measures of standard deviation are very similar between HV and LV agents.} In 4 of our 6 valuation-session combinations is the variance for the LV agent higher than the HV agent in WB. In 6 of our 6 valuation-session combinations is the variance on the HV type greater than the LV type. Permutation tests show this result is signifiant a the 10-\% level.\footnote{If we collapse to session, in 3 of the 4 WB auctions is the variance higher for WB and 4 of the 4 sessions is the variance higher for the LV auction. The value is significant at the 10-\% level for a permutations test.}  At the subject level, 72 of 132 subjects exhibit higher variance as a LV agent than as a HV agent in WB. 71 subjects of 118 exhibit higher variance in the opposite direction in the LV auction. The difference is statistically significant at the 5-\% level.

We now are beginning to unravel the reasons behind mechanism bias. On average, bids in WB are 1.5 equity surpluses lower than LB (note that this is consistent with empirical equilibrium predictions, for at least an agent plays a mixed strategy in equilibria that sustain interior outcomes in the Nash range). These bids are associated with more than a full equity surplus capture for the favored party. If we focus only on deterministic bids, in outcomes that are ultimately efficient (regression (2)), the deterministic bid in WB is 0.168, the deterministic bid in LB is 0.882 (these numbers are statistically different, $p<0.001$). This would correspond with roughly 80\% of the surplus captured by the favored party in each auction (note that this is consistent with Theorems~\ref{Thm:characterization-ee-WB} and~\ref{Thm:characterization-ee-LB}).

In stark contrast is AB. Both predicted LV and HV bids fall within the Nash range of bids (between 0 and 1). This necessarily means the transfer between parties will be in the Nash range. The average value of the two predicted bids is 0.386, which is statistically different than 0.5 ($p<0.01$), an equal split of the surplus, but not particularly far from it. (Recall from Table \ref{tab:payoffs} that the HV bidder earns roughly 17\% more ($p<0.05$) of the surplus than the LV bidder in efficient allocations in AB).

\begin{table}[t]\centering
\begin{tabular}{lccc} \toprule
 & (1) & (2) & (3) \\
&\multicolumn{3}{c}{\begin{tabular}{c}standardized bids\\(equity surpluses over LV maximin)\end{tabular}}\\
& all outcomes & efficient & not efficient\\ \hline
 &  &  &  \\
WB  & -0.627*** & -0.566*** & -0.691*** \\
 & (0.077) & (0.087) & (0.104) \\
LB& 1.089*** & 0.823*** & 1.476*** \\
 & (0.117) & (0.084) & (0.230) \\
HV bidder & 0.264*** & 0.714*** & -0.876*** \\
 & (0.037) & (0.072) & (0.104) \\
WB$\times$& 0.051 & -0.039 & 0.028 \\
HV bidder& (0.055) & (0.088) & (0.164) \\
LB$\times$& 0.079 & 0.593*** & -0.612** \\
HV bidder& (0.091) & (0.183) & (0.252) \\
Constant & 0.245*** & 0.059 & 0.717*** \\
 & (0.056) & (0.067) & (0.066) \\
 &  &  &  \\
\hline
Observations & 9,840 & 7,024 & 2,816 \\
 R-squared & 0.241 & 0.357 & 0.317 \\ \hline
\multicolumn{4}{c}{ *** p$<$0.01, ** p$<$0.05, * p$<$0.1} \\
\end{tabular}
\caption{Regressions of standardized payoffs on auction and bidder type for all allocations, and efficient/inefficient allocations. Standard errors are clustered at the subject level.\label{bidregs}}
\end{table}

\subsection{Weak payoff monotonicity}\label{Sec:payoff-monotonicity}

The distribution of bids provide an explanation as to how the mechanism bias within extreme common value auctions will occur. What we have not explained is why such bid distributions occur. Our theoretical benchmark shows that if empirical distributions are weakly payoff monotone they need to exhibit these patterns if they approach a Nash equilibrium. We now examine whether empirically observed patterns are consistent with weak payoff monotonicity.

A first observation is that the size of our data set is not enough to make full hypothesis testing on weak payoff-monotonicity. 
In our environment with hundreds of actions no realistic data set can test this property. With 451 bidding actions, monotonicity makes directional predictions for 101,926, $(n+1)n/2$ binary relations.  Moreover, to estimate the expected payoff of an action one needs estimates of the whole distribution of play. Thus, one cannot simply reduce the scope of the analysis to some key binary comparisons. Because of this we concentrate on determining whether there is a positive association between empirical expected payoffs and the probability with which agents choose their actions, a feasible task.

\begin{figure}
\centering
\includegraphics[width=0.9\textwidth]{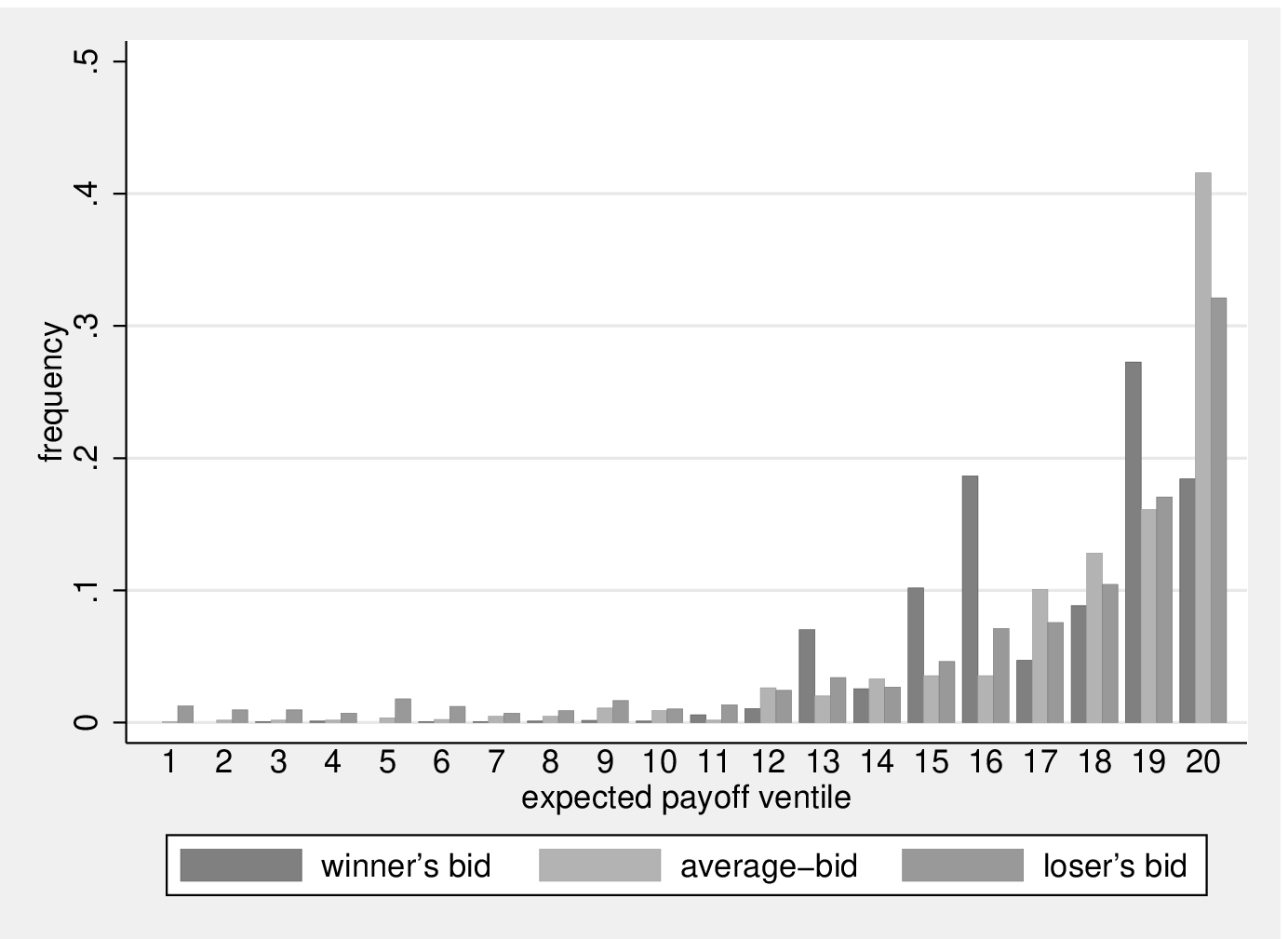}
\includegraphics[width=0.9\textwidth]{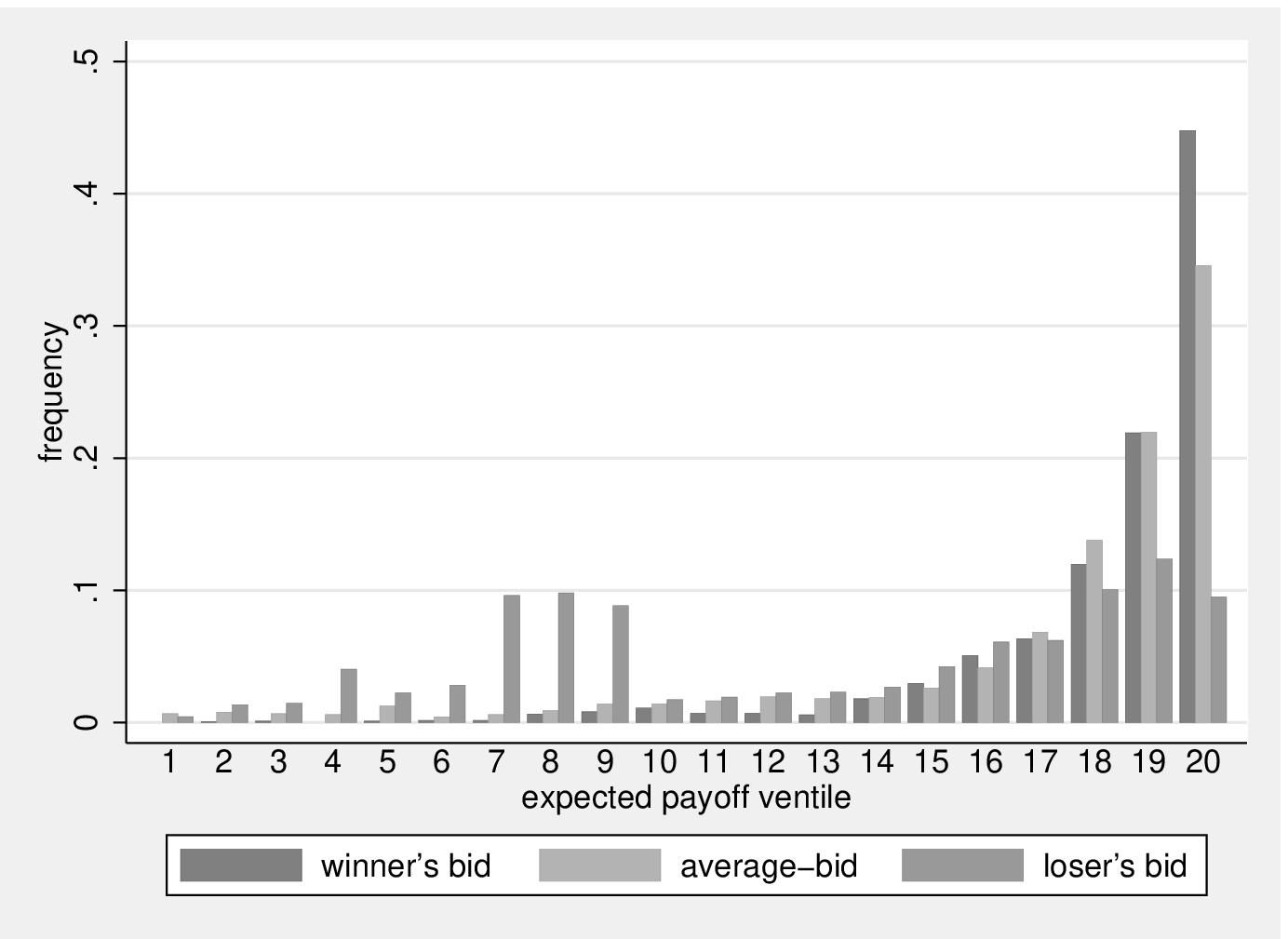}
\caption{\label{Fig:mon1}Aggregate frequency of bids in corresponding ventiles of normalized empirical expected payoffs for (a. top) LV agent (b, bottom) HV agent.}
\end{figure}

As a first approach, which allows us to graphically asses the monotonicity properties of empirical distributions, we calculate the empirical expected payoffs of each bid for both LV and HV agents in each period within each valuation-session. For every period, in each experimental session, we calculate the expected payoffs of making a particular bid as a HV (or LV) against the known distribution of subject bids of other LV (or HV subjects). (Because the experiment featured random matching, there was a $1/(N/2)$ chance any subject would encounter the other $N/2$ subjects of the opposite type. Thus a period's distribution are the ex-ante payoffs a subject would encounter before being matched to a specific player.) We calculate these distributions for each type player, for every bid choice, for every period in our 12 experimental sessions.\footnote{For our purposes, there are not meaningful differences with any of the following analysis if we take the entire valuation-session distribution and ignore period distinctions. In general results look less noisy and provide added emphasis to our main results.}

Following our work in \cite{{Velez-Brown-2019-SP}}, we perform a simple test of a basic property of weak monotonicity: strategies played should have higher payoffs than those that are not played. We focus on the 12 separate sessions or 18 separate valuation-sessions as our level of observation. In both cases all observations have higher average payoffs for strategies played than not played. A sign test rejects the hypothesis that it the difference between these two groups has zero median $(p<0.001)$. Indeed, \cite{{Velez-Brown-2019-SP}} find a similar result across a wide variety of dominant strategy games.

To further illustrate the extent subject play is tied to future payoffs, we rank each possible bid by its expected payoff in each period and sort them into ventiles. Figure~\ref{Fig:mon1} shows a historgram of the expected payoff ventile for the bids chosen by subjects in each of the auction mechanisms. For instance in WB, the HV agent chose a bid associated with the top 5\% of current period expected payoffs in roughly 45\% of all choices. The result is quite astounding; subjects play remarkably well against a future distribution of payoffs.
\begin{table}[t!]\centering
\begin{tabular}{lccc} \hline
 & (1) & (2) & (3) \\
&\multicolumn{3}{c}{bid was selected} \\ \hline
 &  &  &  \\
expected payoff & 0.042*** & 0.028*** & 0.037*** \\
& (0.002) & (0.003) & (0.004) \\
WB $\times$&  &  & 0.002*** \\
expected payoff&  &  & (0.003) \\
LB $\times$&  &  & -0.013*** \\
expected payoff&  &  & (0.003) \\
HV bidder$\times$&  &  & -0.003*** \\
expected payoff&  &  & (0.002) \\
period$\times$&  & 0.001*** & 0.001*** \\
expected payoff&  & (0.000) & (0.000) \\
bid is divisible &&  & 1.752*** \\
by 5 &  &  & (0.090) \\
bid is divisible & &  & 0.763*** \\
by 10 &  &  & (0.074) \\
bid is divisible &  & & 1.112*** \\
by 50 &  &  & (0.080) \\
 &  &  &  \\
\hline
valuation dummies &N &N &Y\\
\hline
subject clusters & 246 & 246 &246\\
observations & 3,301,358 & 3,301,358 & 3,301,358 \\
pseudo r-squared & 0.140 & 0.142 & 0.312 \\ \hline
\multicolumn{4}{c}{ *** p$<$0.01, ** p$<$0.05, * p$<$0.1} \\
\end{tabular}
\caption{Conditional logit regression of bid number on current period expected payoff. Specification 2 adds interaction with period. Specification 3 adds interaction with period, auction type, and bidder type as well as dummy variables for round numbers and valuation.\label{clogit}}
\end{table}

The standard parametric empirical framework to examine the correlation between expected payoffs and decisions is a conditional logistic regression. We follow that framework here. Specifically, we examine over the range of all possible bids whether individuals play higher expected-payoff bids with higher probability. That is,
\begin{equation}
\text{logit}\left(\text{Pr}\left(\text{b}_{ijt}=1|E_t\left(\pi(j)\right)\right)\right)=\beta_1\times E_t\left(\pi(j)\right)+\beta_2\times x_{it}.
\end{equation}
where $b_{ijt}$ is a binary variable denoting whether subject $i$ chose bid $j$ in period $t$. Here, $E_t\left(\pi(j)\right)$
represents the expected payoff (knowing the decision of the other $N/2$ subjects in period $t$ will be selected uniformly at random, see above) of bid choice $j$ in period $t$. Payoff units are in experimental currency units.\footnote{Recall, that one of forty periods was selected at random and converted to cash at 1 point = \$0.10 (The first valuation structure used 1 point = \$0.13). So a single point has expected value of roughly 0.25 of a cent.}

Table \ref{clogit} provides the results of this regression specification. It is clear that the probability of a strategy played is positively correlated with its expected value in the current period. Average marginal effects (not shown), suggest a gain in expectation of 100 experimental points increases the likelihood of that bid being chosen by a subject by 15 probability points (from specification 1). Payoff responsiveness of subjects increases over time (see period interaction variable in specifications (2) and (3)), though the marginal effects of those specifications suggest these results do not have economic significance. This effect is robust to controlling for session type, experience, and whether the bid is a round number, all of which are also predictive of play.

Our theoretical model relied on the assumption of monotonicity to show in this environment only certain equilibria would be possible under certain mechanisms. Our analysis to this point has coincided with the theory in lockstep. First, subjects obey a loose definition of monotonicity, as they play strategies associated with a expected higher payoff more often. Second, bids tend to be near the low end of the Nash distribution in WB, the high end of the Nash distribution in LB and the middle range of the distribution in AB. This directly means that in efficient outcomes a bias is present; WB, than AB, than LB are best for the party with the HV on the item. Because efficient outcomes occur more often than not (shown next section) this translates to overall mechanism advantages following the same order.

\begin{table}[t]
\centering
\resizebox{0.9\textwidth}{!} {
\begin{tabular}{lccccccc}
\toprule
&\multicolumn{3}{c}{efficient outcomes}&&\multicolumn{3}{c}{equilibrium outcomes}\\
\begin{tabular}{c}valuation\\structure\end{tabular}& WB & AP &LB& &WB& AP & LB\\
\midrule
\begin{tabular}{c}1A\\(20 periods)\end{tabular}& \begin{tabular}{c}0.720\\(0.450)\\200\end{tabular}&\begin{tabular}{c}0.693\\(0.462)\\300\end{tabular}&\begin{tabular}{c}0.650\\(0.478)\\220\end{tabular}&&\begin{tabular}{c}0.690\\(0.464)\\200\\\end{tabular}&\begin{tabular}{c}0.630\\(0.484)\\300\end{tabular}&\begin{tabular}{c}0.518\\(0.501)\\220\end{tabular}\\
\vfivefive
\begin{tabular}{c}1B\\(20 periods)\end{tabular}& \begin{tabular}{c}0.920\\(0.272)\\200\end{tabular}&\begin{tabular}{c}0.780\\(0.415)\\300\end{tabular}&\begin{tabular}{c}0.755\\(0.431)\\220\end{tabular}&&\begin{tabular}{c}0.915\\(0.280)\\200\end{tabular}&\begin{tabular}{c}0.770\\(0.422)\\300\end{tabular}&\begin{tabular}{c}0.709\\(0.455)\\220\end{tabular}\\
\vfivefive
\begin{tabular}{c}2A\\(20 periods)\end{tabular}& \begin{tabular}{c}0.658\\(0.475)\\260\end{tabular}&\begin{tabular}{c}0.606\\(0.490)\\180\end{tabular}&\begin{tabular}{c}0.572\\(0.496)\\180\end{tabular}&&\begin{tabular}{c}0.446\\(0.498)\\260\end{tabular}&\begin{tabular}{c}0.428\\(0.496)\\180\end{tabular}&\begin{tabular}{c}0.250\\(0.434)\\180\end{tabular}\\
\vfivefive
\begin{tabular}{c}2B\\(20 periods)\end{tabular}& \begin{tabular}{c}0.831\\(0.376)\\260\end{tabular}&\begin{tabular}{c}0.822\\(0.383)\\180\end{tabular}&\begin{tabular}{c}0.689\\(0.464)\\180\end{tabular}&&\begin{tabular}{c}0.827\\(0.379)\\260\end{tabular}&\begin{tabular}{c}0.806\\(0.397)\\180\end{tabular}&\begin{tabular}{c}0.633\\(0.483)\\180\end{tabular}\\
\vfivefive
\begin{tabular}{c}3\\(40 periods)\end{tabular}& \begin{tabular}{c}0.818\\(0.387)\\400\end{tabular}&\begin{tabular}{c}0.772\\(0.420)\\400\end{tabular}&\begin{tabular}{c}0.685\\(0.465)\\400\end{tabular}&&\begin{tabular}{c}0.745\\(0.436)\\400\end{tabular}&\begin{tabular}{c}0.680\\(0.467)\\400\end{tabular}&\begin{tabular}{c}0.490\\(0.501)\\400\end{tabular}\\
\vfivefive
\begin{tabular}{c}4\\(40 periods)\end{tabular}& \begin{tabular}{c}0.680\\(0.467)\\400\end{tabular}&\begin{tabular}{c}0.600\\(0.491)\\280\end{tabular}&\begin{tabular}{c}0.589\\(0.493)\\360\end{tabular}&&\begin{tabular}{c}0.367\\(0.483)\\400\end{tabular}&\begin{tabular}{c}0.243\\(0.430)\\280\end{tabular}&\begin{tabular}{c}0.258\\(0.438)\\360\end{tabular}\\
\midrule
overall& \begin{tabular}{c}0.764\\(0.425)\\1,720\end{tabular}&\begin{tabular}{c}0.717\\(0.451)\\1,640\end{tabular}&\begin{tabular}{c}0.655\\(0.475)\\1,560\end{tabular}&&\begin{tabular}{c}0.638\\(0.481)\\1,720\end{tabular}&\begin{tabular}{c}0.599\\(0.490)\\1,640\end{tabular}&\begin{tabular}{c}0.460\\(0.499)\\1,560\end{tabular}\\
\bottomrule
\end{tabular}
}
\caption{\label{tab:outcomes} Mean, standard deviation, and number of observations of standardized bids, by auction type, session$\times$valuation structure for HV and LV agents.}
\end{table}

\subsection{Efficient and equal-income competitive equilibrium outcomes}\label{Sec:efficiency}
The purpose of the auctions mechanisms in this paper is to achieve efficient outcomes for both players with transfers that fall under prices that could conceivably observed in a market with equal incomes. To that end, it is important to examine how often these mechanism worked as intended. That is, how often the bidder with the high value on item B received that item, and how often this was accompanied by a transfer that is associated with an equilibrium outcome. Note that these outcomes occur at the subject-pair level, as in every period of this experiment, two subjects observed the exact same outcome.
%

As before we first observe session averages. Table \ref{tab:outcomes} provides valuation-session averages of efficient and equilibrium outcomes in each of the twelve experimental sessions. For each measure, in each session-valuation pair, WB achieves the highest percentage of desired outcomes, then AP, then  LB. Admittedly, this relationship was not hypothesized by any theory. However, even if we allow an additional degree of freedom because we may not necessarily expect this relationship, permutation tests would find the ordering significant at the 1-\% level.\footnote{If we think of each session as independent and do not count the first session, the next three sessions are the most extreme of 216 ($6^3$) observations ($p<0.01$) to reject the hypothesis.} Table \ref{outcomereg} shows the results of a regression using additional controls, jointly clustered on subject and type (who is HV and LV) and achieves nearly identical results.

It is important to think about the implications of this result, even though it was not necessarily predicted by theory ex-ante. On the one hand, our quantitative exercise in Sec.~\ref{Sec:model} revealed that the relative performance of these auctions can be affected by the details of the environment. On the other hand, for the particular model for which the quantification was done, moving towards best responses increases the efficiency of the all mechanisms in the different environments in which calculations were done. In view of this computational benchmark, the result is not particularly surprising.  WB is a mechanism unlike the other two. It is plausible that subjects entering a laboratory experiment would already have some experience with the first-price auction, a related popular auction format.

\begin{table}[t]\centering
\begin{tabular}{lcc} \toprule
& efficient outcome & equilibrium outcome \\ \midrule
 &  &  \\
WB  & 0.061** & 0.082*** \\
 & (0.024) & (0.031) \\
LB & -0.054* & -0.114*** \\
 & (0.029) & (0.035) \\
Period within valuation & 0.004*** & 0.007*** \\
 & (0.001) & (0.001) \\
120-320 valuation & 0.145*** & 0.352*** \\
 & (0.026) & (0.031) \\
250-450 valuation & 0.095*** & 0.337*** \\
 & (0.029) & (0.034) \\
250-290 valuation & 0.219*** & 0.516*** \\
 & (0.032) & (0.036) \\
Constant & 0.549*** & 0.211*** \\
 & (0.026) & (0.036) \\
 &  &  \\ \midrule
Observations & 4,920 & 4,920 \\
R-squared & 0.044 & 0.175 \\ \bottomrule
\multicolumn{3}{c}{ *** p$<$0.01, ** p$<$0.05, * p$<$0.1} \\
\end{tabular}
\caption{Regressions of efficient and equilibrium outcomes on auction, valuation, and period within valuation. Standard errors are 2-way clustered on each subject$\times$bidder-type pair.\label{outcomereg}}
\end{table}

\section{Discussion and concluding remarks}\label{Sec:Discussion}

We have shown experimental evidence supporting that WB and LB are biased mechanisms and that AB balances better the interests of both agents in complete information partnership dissolution environments. WB benefits the HV agent and LB benefits the LV agent. AB has lower efficiency than WB in our experiments, however. The efficiency ranking seems to be connected with the degree to which agents understand the mechanisms. Thus, one can conclude that an arbitrator who selects a mechanism between these three is unambiguously better off by not using LB. It would be interesting to see if the performance of AB can increase with alternative ways in which this mechanism is framed and explained to individuals. In this case a social planner may not have to choose between efficiency and equity when selecting a mechanism to dissolve a partnership.

We conclude with a discussion of alternative refinements of Nash equilibrium for the WB and LB games.

A substantial literature in game theory has studied the plausibility of equilibria \citep[see][for an early survey]{VanDamme-1991-Springer}. In general, the refinements proposed in these studies are founded on decision theoretical arguments that implicitly or explicitly assume ``admissibility,'' i.e., weakly dominated behavior is not plausible.

In our environment, any refinement satisfying admissibility selects a single equilibrium outcome for each of WB and LB. The HV agent selects a standardized bid of 0 in WB and the LV agent selects a standardized bid of 1 in LB (these are the payoff determinant bids). Clearly, these equilibrium refinements also predict the direction of mechanism bias revealed by empirical equilibrium analysis.

In general, refinements satisfying admissibility are not consistently good predictors at selecting an equilibrium from empirical data. A variety of experiments with dominant strategy mechanisms, with a unique dominant strategy equilibrium, have shown the prevalence of weakly dominated behavior in these games \citep[see][for a meta-study of these papers]{Velez-Brown-2019-SP}. Moreover, \citet{Goeree-Holt-Palfrey-2016-Book} note several instances where logit QRE converges to a dominated equilibrium. In all these cases, logit QRE is the better predictor compared, for instance, with trembling-hand-perfect equilibrium. (Recall that Logistic QRE satisfy weak payoff monotonicity).

\begin{figure}[t]
\centering
\includegraphics[width=0.75\textwidth]{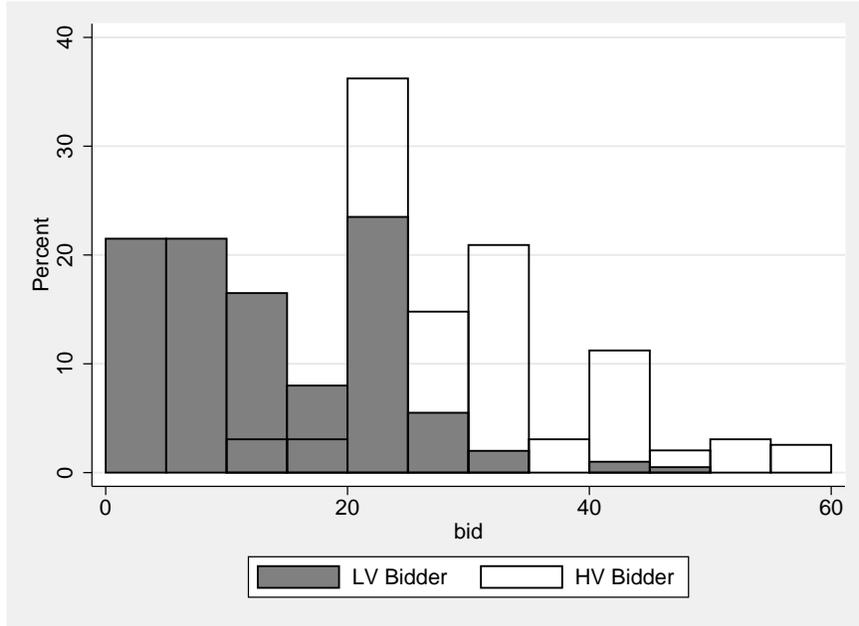}
\caption{\textbf{Distribution of bids WB session type 1B}. The Nash range for this valuation structure is $\{10,...,110\}$. All bids of the LV agent and 98\% of the HV agent fall in the range shown in the figure.}\label{Fig:Distribution-bid-1B}
\end{figure}

Weakly dominated actions may be compelling in an extreme-price auction \citep[see Sec. 6.4][for a related discussion]{Velez-Brown-2018-EE}. Consider for instance our valuation structure 1B, in which the Nash range is $\{10,...,110\}$. Let $b_l>10$. One can easily see that $b_l$ is weakly dominated by $10$ for the LV agent in WB. Utility maximization predicts that the LV agent bids~$b_l$ only if agent~$h$ never bids to the left of $b_l$. Thus, it is tempting to think that the LV agent, considering a small deviation by the HV agent will always preemptively bid $10$ instead. If the LV agent follows this safe choice consistently, the incentive of the HV agent is to bid also $10$, leaving the LV agent with the lowest possible share of the equity surplus.

One can argue that bidding $b_l$ above $10$ is normatively compelling and intuitively plausible, however. This is the only way that the LV agent can enforce a more balanced division of the equity surplus. Thus the risk that the LV agent takes when bidding $10$ is the means to enforce a better equilibrium. Moreover, when $b_l$ is still close to $10$, the LV agent is practically bearing no risk, for this agent would lose very little when the HV agent bids below $b_l$. If $b_l$ satisfies the conditions of our Theorem~\ref{Thm:characterization-ee-WB}, the LV agent can bid $b_l$ with enough probability to keep in check the incentives of the HV agent and still satisfy weak payoff monotonicity. The HV agent on the other hand, would lose much more than the LV agent in case this last agent ends up getting the object and paying $b_l$ for it. Thus, the probability that the HV agent bids below $10$ should also be very small if the LV agent is consistently biding at least $b_l$ with (enough) positive probability.  Thus, not only the LV agent may not care much about the risk of buying for $b_l$, but also if the HV agent is taking notice of this behavior, the probability that this happens is also very small, reinforcing the incentive of the LV agent to bid~$b_l$. Thus, one can make the case that it is plausible to observe bids accumulating in the interior of the Nash range. Data from our experiment for WB with this valuation structure seems in line with the intuition that we just described (Fig.~\ref{Fig:Distribution-bid-1B}).\footnote{This was the maximal efficiency session in our experiment (92\%)). This means that this is the session in which bid distributions of LV and HV agents separated the most. This makes this session our best example to illustrate this phenomenon.}

Thus, imagine that an arbitrator operates WB in a society in which agents' valuations are correlated with a characteristic that is protected by law against statistical discrimination. For instance suppose that the population can be partitioned in two sets, $H$ and $L$, and that on average agents in $H$ hold higher valuations than agents in $L$. Suppose that $L$ preset a class action suit against the arbitrator and argue, based on a theoretical benchmark and experimental data corroborating it, that this mechanism is systematically biased against them (field data with observable valuations is likely to be unfeasible). If the theoretical benchmark is Nash equilibrium, the argument is easily dismissed based on the multiplicity of equilibria that do not exhibit the bias. If the theoretical benchmark is trembling-hand-perfection or other equilibrium refinement that dismisses all weakly dominated behavior, the argument is easily dismissed based on the long history of experiments in which weakly dominated behavior is persistently observed. For instance, our Fig.~\ref{Fig:Distribution-bid-1B} would be a good Exhibit A. A spin doctor would argue that since LV agents in this session-valuation (or any of the aforementioned experiments) are able to hold their ground against HV agents and keep them from bidding below 20 units, it is plausible that they can do so for any amount in the Nash range. By contrast, an argument based on empirical equilibrium analysis is not easily dismissed. Indeed, to do so one needs to dismiss the tendency of agents to select actions that give them higher payoffs. If this property is satisfied in data, empirical distributions of WB will look like Fig.~\ref{Fig:Distribution-bid-1B} as bids become separated and efficiency is achieved. It is possible that LV holds their ground and bid more than $10$, say $b_l$. There is a limit, however. If the agents' actions are informed by expected utility, as $b_l$ increases, there is more and more actions that are better than $b_l$ for $LV$. If these actions are played with sufficient probability, comparatively speaking with respect to the probability with which LV plays $b_l$, then this bid cannot be too far from $10$. This is indeed the intuition captured by empirical equilibrium, which is confirmed in our data.


\bibliography{ref-EBEX}

\newpage
\section*{Appendix not for publication}

We denote the strategy that selects bid $b$ for sure by $\delta_b$. The expected payoff of agent $i$ from bidding $b$ in the $\alpha$-auction when the other agent plays strategy $\sigma_{-i}$  is  $U_\alpha(\delta_{b}|\sigma_{-i};v_i)$.

\begin{proof}[\textbf{Proof of Theorem~\ref{Thm:characterization-ee-WB}}]Let $\sigma$, $\{\sigma^\lambda\}$, and $t(v)$  be as in the statement of the lemma. By Proposition~\ref{Prop:charc-equili-auctions} there is $p\in\{c_l,...,c_h\}$ in the support of $\sigma_l$ and $\sigma_h$ such that the support of $\sigma_l$ belongs to $\{0,...,p\}$ and the support of $\sigma_h$ belongs to $\{0,...,\overline{p}\}$. If $p=c_l$, statements 1-3 follow from convergence of $\sigma^\lambda$, i.e., because as $\lambda\rightarrow\infty$, $\sigma^\lambda\rightarrow\sigma$; statement 4 holds because $\pi_l(\sigma)=c_l$. Suppose that $p=c_l+1$.  Then, $ES(v)\geq 1$, and $t(v)>1$. Statement 2 follows from convergence of $\sigma^\lambda$. Statement 3 follows from convergence given that $\pi_l(\sigma)=c_l+1$ and $\pi_h(\sigma)=c_h+(ES(v)-1)$. Recall that $v_h>v_l\geq 0$. If $v_l\geq v_h/3$, then $v_l>0$. Then,
\[U_1(\delta_{c_l-1}|\sigma^\lambda_h;v_l)-U_1(\delta_{p}|\sigma^\lambda_h;v_l)=\sum_{r<c_l-1}\sigma_h^\lambda(r)2+\sigma_h^\lambda(c_l)(1)>0,\]
and
\[U_1(\delta_{c_l}|\sigma^\lambda_h;v_l)-U_1(\delta_{p}|\sigma^\lambda_h;v_l)=\sum_{r<c_l}\sigma_h^\lambda(r)+\sigma_h^\lambda(c_l)(1)>0.\]
By monotonicity, $\sigma^\lambda_l(c_l-1)\geq\sigma^\lambda_l(p)$ and $\sigma^\lambda_l(c_l)\geq\sigma^\lambda_l(p)$. By convergence, $\sigma_l(c_l-1)\geq\sigma_l(p)$ and $\sigma_l(c_l)\geq\sigma_l(p)$. Since the support of $\sigma_l$ belongs to $\{0,...,p\}$, then $E_{\sigma_l}(b)\leq c_l$. Statement 1 follows from convergence of $\sigma^\lambda$. Since $E_{\sigma_h}(b)=c_l+1$, statement 4 follows from convergence of $\sigma^\lambda$.

Suppose that $\tau\equiv p-c_l>1$. Let $y\equiv\max\{0,c_l-(3\tau-1)\}$. An argument as in the proof of Theorem 1 in \cite{Velez-Brown-2019-EBA} shows that there is $\Lambda\in \N$ such that for each $\lambda\geq \Lambda$ and each $y\leq b\leq p$,
\begin{equation}\sigma^\lambda_l(b)\geq \sigma^\lambda_l(p).\label{Eq:bound1}\end{equation}
Thus,
\begin{equation}\label{Eq:upper-bound-sigma_l}
\sigma_l(p)\leq1/\min\{4\tau,\tau+c_l+1\}.\end{equation}
We claim that $p\leq c_l+ t(v)-1$. Suppose by contradiction that $p-c_l>\max\{2c_h/3-c_l,ES(v)/3\}$.   Since $\sigma$ is a Nash equilibrium, $U_1(\delta_p|\sigma_l;v_h)\geq U_1(\delta_{p-1}|\sigma_l;v_h)$. Thus,
\[v_h-p\geq(1-\sigma_l(p))(v_h-p+1)+\sigma_l(p)(p).\]
Equivalently,
\begin{equation}\sigma_l(p)\geq1/(2(c_h-c_l)-2\tau+1).\label{Eq:Equil-sigma_l(p)}\end{equation}
By (\ref{Eq:upper-bound-sigma_l}) and (\ref{Eq:Equil-sigma_l(p)}),
\[2(c_h-c_l)-2t+1\geq \min\{4\tau,\tau+c_l+1\}.\]
Suppose that $\min\{4\tau,\tau+c_l+1\}=4\tau$. Then,
\[(c_h-c_l)/3+1/6=ES(v)/3+1/6\geq \tau=p-c_l.\]
Since $c_h-c_l$ and $p-c_l$ are integers, $(c_h-c_l)/3=ES(v)\geq p-c_l$. This is a contradiction. Suppose then that $\min\{4\tau,\tau+c_l+1\}=\tau+c_l+1$. Then,
\[2c_h/3-c_l\geq \tau=p-c_l.\]
This is a contradiction.

Since $E_{\sigma_h}(b)=p<c_l+ t(v)-1/2$, $\pi_l(\sigma)=c_l+p$, and $\pi_h(\sigma)=c_h+(ES(v)-p)$,  statements 2 and 3 follow from convergence of $\sigma^\lambda$. Since the support of $\sigma_l$ belongs to $\{0,...,p\}$, by (\ref{Eq:bound1}), $E_{\sigma_l}(b)\leq (p+c_l)/2<p=E_{\sigma_h}(b)$. Then, statement $4$ follows from convergence of $\sigma^\lambda$.

Finally, suppose that $v_l\geq v_h/3$. We claim that $c_l-(p-c_l)\geq0$.  Consider first the case that $v_l\leq v_h/2$. Then, $c_l\leq c_h/2$ and consequently $(c_h-c_l)/3\leq2c_h/3-c_l$. That is, $\max\{2c_h-c_l,ES(v)/3\}=2c_h/3-c_l$. Thus, $p\leq 2c_h/3$. Since $c_l\geq c_h/3$, $p-c_l\leq c_l$. If $v_l>v_h/2$, $\max\{2c_h-c_l,ES(v)/3\}=ES(v)/3$. Thus, $p-c_l\leq ES(v)/3=(c_h-c_l)/3\leq c_h/6<c_l$. An argument as that in the proof of Theorem 1 in \citet{Velez-Brown-2019-EBA}, shows that for each $q>c_l$ such that $c_l-(q-c_l)>0$, $\sigma^\lambda_l(c_l-(q-c_l))>\sigma^\lambda_l(q)$. Thus, $E_{\sigma_l}(b)<c_l$. Thus statement 1 follows from convergence of $\sigma^\lambda$.

\medskip
We earlier claimed that an argument as in the proof of Theorem 1 in \cite{Velez-Brown-2019-EBA} shows that for each $\lambda\in\N$ and each $c_l<q$ such that $c_l-(q-c_l)\geq 0$,
\[\sigma^\lambda_l(c_l-(q-c_l))>\sigma^\lambda_l(q),\]
and that there is $\Lambda\in \N$ such that for each $\lambda\geq \Lambda$, and each $y\leq b\leq p$, $\sigma^\lambda_l(b)\geq \sigma^\lambda_l(p)$. We present now an explicit argument.

For  $i\in N$  and  $\{b,d\}\subseteq\{0,1,...,\overline{p}\}$, let $\Delta_i(b,d)$ be the difference in expected utility for agent $i$ in \textit{WB} between the two situations in which agent $i$ bids strictly to the left of $b$ and bids exactly $d$, conditional on agent $-i$ bidding $b$. Using this notation we have that when $b<d$,
\[\begin{array}{rl}U_1(\delta_b|\sigma_{h}^\lambda,v_l)-U_1(\delta_d|\sigma_{h}^\lambda,v_l)=&
\sum_{r<b}\sigma_{h}^\lambda(r)(d-b)
+\sigma_{h}^\lambda(b)(b-(2c_l-d))
\\&+\sum_{b<r\leq d}\sigma_{h}^\lambda(r)\Delta_l(r,d).\end{array}\]
Let $c_l<q\leq p$, $\rho=q-c_l$, and $c_l-\rho\leq b<q$,
\[\begin{array}{rl}U_1(\delta_b|\sigma_{h}^\lambda,v_l)-U_1(\delta_q|\sigma_{h}^\lambda,v_l)=&
\sum_{r<b}\sigma_{h}^\lambda(r)(q-b)
+\sigma_{h}^\lambda(b)(b-(2c_l-q))
\\&+\sum_{b<r\leq q}\sigma_{h}^\lambda(r)\Delta_l(r,q).\end{array}\]
Now, $b-(2c_l-q)=b-2c_l+(c_l+\rho)=b+(c_l-\rho)\geq0$, $\Delta_l(q,q)=0$, and for each $b<r<q$, $\Delta_l(r,q)=\rho-(c_l-r)>0$. Thus,
\[\begin{array}{rl}U_1(\delta_b|\sigma_{h}^\lambda,v_l)-U_1(\delta_q|\sigma_{h}^\lambda,v_l)>0.\end{array}\]
By monotonicity
\[\sigma_l^\lambda(b)>\sigma_l^\lambda(q).\]
This proves that for each $c_l<q$ such that $c_l-(q-c_l)\geq 0$,
\[\sigma^\lambda_l(c_l-(q-c_l))>\sigma^\lambda_l(q),\]
and that for each $c_l-\tau\leq b\leq p$,
\begin{equation}\sigma_l^\lambda(b)>\sigma_l^\lambda(p).\label{Eq:increasing-sigma_llambda1}\end{equation}
We complete the proof by induction on $b$. Let $y<b\leq c_h-\tau$. Suppose that there is $\Lambda\in\N$ such that for each $\lambda\geq \Lambda$ and for each $b\leq r\leq p$, $\sigma_l^\lambda(r)\geq \sigma_l^\lambda(p)$. We prove that $\Lambda$ can be selected large enough so for each $\lambda\geq \Lambda$ and for each $b-1\leq r\leq p$, $\sigma_l^\lambda(r)\geq \sigma_l^\lambda(p)$.   By convergence, for each $b\leq r\leq p$, \begin{equation}\sigma_l(b)\geq \sigma_l(p).\label{Eq:increasing-sigma_l1}\end{equation}
Let $b\leq s<p$. Then,
\[\begin{array}{rl}U_1(\delta_s|\sigma_{l},v_h)-U_1(\delta_{s-1}|\sigma_{l},v_h)=&\sum_{r\leq s-1}\sigma(r)(-1)+\sigma_l(s)(v_h-s-s)\\
&=\sum_{r\geq s}\sigma_l(r)-1+\sigma_l(s)2(c_h-s).
\end{array}\]
By (\ref{Eq:increasing-sigma_l1}) and since $c_h-s>c_h-p=c_h-(c_l+\tau)$,
\[\begin{array}{rl}U_1(\delta_s|\sigma_{l},v_h)-U_1(\delta_{s-1}|\sigma_{l},v_h)>&
2\sigma_l(p)-1+\sigma_l(p)2(c_h-c_l-\tau).
\end{array}\]
By (\ref{Eq:Equil-sigma_l(p)}), $\sigma_l(p)\geq1/(2(c_h-c_l)-2\tau+1)$. Thus,
\[\begin{array}{rl}U_1(\delta_s|\sigma_{l},v_h)-U_1(\delta_{s-1}|\sigma_{l},v_h)>&
\frac{2(c_h-c_l-\tau)+2}{2(c_h-c_l)-2\tau+1}-1>0.
\end{array}\]
Since $\sigma^\lambda\rightarrow\sigma$, there is $\Lambda\in\N$ such that for each $\lambda\geq \Lambda$,
\[\begin{array}{rl}U_1(\delta_s|\sigma_{l}^\lambda,v_h)-U_1(\delta_{s-1}|\sigma_{l}^\lambda,v_h)>0.
\end{array}\]
By monotonicity, for each $b\leq s<p$,
\begin{equation}\label{Eq:decreasing-sigmah}
\sigma_h^\lambda(s-1)\leq \sigma_h^\lambda(s).
\end{equation}
Now,
\[\begin{array}{rl}U_1(\delta_{b-1}|\sigma_{h}^\lambda,v_l)-U_1(\delta_p|\sigma_{h}^\lambda,v_l)=&
\sum_{r<b-1}\sigma_{h}^\lambda(r)(p-b+1)
\\&+\sigma_{h}^\lambda(b-1)(b-1-(2c_l-p))
\\&+\sum_{b-1<r\leq p}\sigma_{h}^\lambda(r)\Delta_l(r,p)\\
>&\sigma_{h}^\lambda(b-1)(b-1-(2c_l-p))
\\&+\sum_{b-1<r\leq p}\sigma_{h}^\lambda(r)\Delta_l(r,p)
\\=&\sum_{b-1\leq r\leq p}\sigma_{h}^\lambda(r)\Delta_l(r,p)
.\end{array}\]
Since for each $r<p$, $\Delta_l(r,p)=\tau-(c_l-r)$,
\[\begin{array}{rl}U_1(\delta_{b-1}|\sigma_{h}^\lambda,v_l)-U_1(\delta_p|\sigma_{h}^\lambda,v_l)>&\sum_{1\leq x\leq c_l-\tau-b+1}(\sigma_l^\lambda(c_l-\tau+x)\\&-\sigma_l^\lambda(c_l-\tau-x))\Delta_l(c_l-\tau+x)
\\&+\sum_{c_l-\tau+(c_l-\tau-b+1)<r\leq p}\sigma_{h}^\lambda(r)\Delta_l(r,p)
\\\geq &0,\end{array}\]
where the last inequality follows from (\ref{Eq:decreasing-sigmah}). By monotonicity, $\sigma^\lambda_l(b-1)\geq\sigma^\lambda_l(p)$.
\end{proof}
\end{document}